\documentclass[a4paper,11pt]{article}
\pdfoutput=1 % if your are submitting a pdflatex (i.e. if you have
\usepackage{jheppub} % for details on the use of the package, please
\usepackage{mathrsfs}
\usepackage{bbm}
\usepackage{bm}
\usepackage{array,longtable}
\usepackage{multirow}
\usepackage{slashed}
\usepackage[utf8]{inputenc}

\allowdisplaybreaks

\title{\boldmath New physics in the angular distribution of $B_c^- \to J/\psi (\to \mu^+ \mu^-)\tau^- (\to \pi^- \nu_\tau)\bar{\nu}_\tau$ decay}

\author[a]{Quan-Yi Hu,}
\author[b]{Xin-Qiang Li,}
\author[b]{Xiao-Long Mu,}
\author[b]{Ya-Dong Yang}
\author[b]{and Dong-Hui Zheng}

\affiliation[a]{School of Physics and Electrical Engineering, Anyang Normal University,
\\Anyang, Henan 455000, China}
\affiliation[b]{Institute of Particle Physics and Key Laboratory of Quark and Lepton Physics~(MOE), 
\\Central China Normal University, Wuhan, Hubei 430079, China}

\emailAdd{qyhu@aynu.edu.cn}
\emailAdd{xqli@mail.ccnu.edu.cn}
\emailAdd{muxl@mails.ccnu.edu.cn}
\emailAdd{yangyd@mail.ccnu.edu.cn}
\emailAdd{zhengdh@mails.ccnu.edu.cn}

\abstract{In $B_c^- \to J/\psi (\to \mu^+ \mu^-)\tau^-\bar{\nu}_\tau$ decay, the three-momentum $\boldsymbol{p}_{\tau^-}$ cannot be determined accurately due to the decay products of $\tau^-$ inevitably include an undetected $\nu_{\tau}$. As a consequence, the angular distribution of this decay cannot be measured. In this work, we construct a {\it measurable} angular distribution by considering the subsequent decay $\tau^- \to \pi^- \nu_\tau$. The full cascade decay is  $B_c^- \to J/\psi (\to \mu^+ \mu^-)\tau^- (\to \pi^- \nu_\tau)\bar{\nu}_\tau$, in which the three-momenta $\boldsymbol{p}_{\mu^+}$, $\boldsymbol{p}_{\mu^-}$, and $\boldsymbol{p}_{\pi^-}$ can be measured. The five-fold differential angular distribution containing all Lorentz structures of the new physics (NP) effective operators can be written in terms of twelve angular observables $\mathcal{I}_i (q^2, E_\pi)$. Integrating over the energy of pion $E_\pi$, we construct twelve normalized angular observables $\widehat{\mathcal{I}}_i(q^2)$ and two lepton-flavor-universality ratios $R(P_{L,T}^{J/\psi})(q^2)$. Based on the $B_c \to J/\psi$ form factors calculated by the latest lattice QCD and sum rule, we predict the $q^2$ distribution of all $\widehat{\mathcal{I}}_i$ and $R(P_{L,T}^{J/\psi})$ both within the Standard Model and in eight NP benchmark points. We find that the benchmark BP2 (corresponding to the hypothesis of tensor operator) has the greatest effect on all $\widehat{\mathcal{I}}_{i}$ and $R(P_{L,T}^{J/\psi})$, except $\widehat{\mathcal{I}}_{5}$. The ratios $R(P_{L,T}^{J/\psi})$ are more sensitive to the NP with pseudo-scalar operators than the $\widehat{\mathcal{I}}_{i}$. Finally, we discuss the symmetries in the angular observables and present a model-independent method to determine the existence of tensor operators.}

\begin{document} 
\maketitle
\flushbottom

\section{Introduction}
\label{sec:introduction}
Exploring new physics (NP) beyond the Standard Model (SM) has been one of the most important tasks in high energy physics, especially since the discovery of Higgs boson~\cite{Aad:2012tfa,Chatrchyan:2012ufa,Aad:2015zhl}. In recent years, the existence of NP that breaks the universality of lepton flavour in $b \to c \tau^- \bar{\nu}_\tau$ transition has been implied by the anomalous measurements~\cite{Lees:2012xj,Lees:2013uzd,Aaij:2015yra,Huschle:2015rga,Hirose:2016wfn,Aaij:2017uff,Hirose:2017dxl,Aaij:2017deq,Belle:2019rba} on $\bar{B} \to D^{(*)} \tau^- \bar{\nu}_\tau$ decays. Moreover, the averaging results performed by the Heavy Flavor Averaging Group (HFLAV)~\cite{Amhis:2019ckw} show that the measurements of $ R(D^{(*)})\equiv\mathcal{B}(\bar{B}\to D^{(*)} \tau\bar \nu)/\mathcal{B}(\bar{B}\to D^{(*)} \ell\bar \nu)$ ($\ell=e,\,\mu$) deviate about $3.1 \sigma$~\cite{Amhis:2019up} from the predicted values within the SM.\footnote{The recent ref.~\cite{Bordone:2019vic} finds $ R(D^{*})=0.250\pm 0.003$ in the SM. Since the HFLAV average does not include this value, the actual deviation from the SM prediction is larger than $3.1 \sigma$. Very recently, the review~\cite{Bernlochner:2021vlv} revisits the $R(D^{(*)})$ world averages by investigating the correlations between the systematic uncertainties of all measurements, and shows that their averages lead to an increased tension of about $3.6\sigma$ with respect to the SM.} Motivated by this deviation and using the available data on $b \to c \tau^- \bar{\nu}_\tau$ transition, a number of global fitting analyses have been carried out~\cite{Alok:2017qsi,Hu:2018veh,Alok:2019uqc,Murgui:2019czp,Blanke:2018yud,Blanke:2019qrx,Shi:2019gxi,Cheung:2020sbq,Kumbhakar:2020jdz}, finding that some different combinations of NP coupling parameters can well explain the $R(D^{(*)})$ anomaly. At the same time, a large number of works have been completed in some specific NP models, such as  leptoquarks~\cite{Tanaka:2012nw,Sakaki:2013bfa,Bauer:2015knc,Fajfer:2015ycq,Li:2016vvp,Crivellin:2017zlb,Becirevic:2018afm,Angelescu:2018tyl,Bansal:2018nwp,Iguro:2018vqb,Crivellin:2019dwb}, $R$-parity violating supersymetric models~\cite{Deshpand:2016cpw,Altmannshofer:2017poe,Hu:2018lmk,Hu:2020yvs,Altmannshofer:2020axr}, charged Higgses~\cite{Tanaka:2012nw,Crivellin:2012ye,Celis:2012dk,Ko:2012sv,Celis:2016azn,Iguro:2017ysu,Iguro:2018qzf}, charged vector bosons~\cite{Asadi:2018wea,Greljo:2018ogz,Babu:2018vrl}, and Pati-Salam Model~\cite{Blanke:2018sro,Iguro:2021kdw}. These NP effects also affect other $b \to c $ semileptonic decay modes, such as $B_c \to J/\psi \tau \bar{\nu}_\tau$~\cite{Aaij:2017tyk,Harrison:2020nrv,Harrison:2020gvo,Watanabe:2017mip,Wei:2018vmk,Tran:2018kuv,Issadykov:2018myx,Cohen:2018vhw,Cohen:2018dgz,Huang:2018nnq,Wang:2018duy,Leljak:2019eyw,Hu:2019qcn,Azizi:2019aaf,Penalva:2020ftd}, $B_c \to \eta_c \tau \bar{\nu}_\tau$~\cite{Wei:2018vmk,Tran:2018kuv,Issadykov:2018myx,Berns:2018vpl,Murphy:2018sqg,Wang:2018duy,Leljak:2019eyw,Hu:2019qcn,Azizi:2019aaf,Penalva:2020ftd}, $\Lambda_b \to \Lambda_c \tau \bar{\nu}_\tau$~\cite{Dutta:2015ueb,Li:2016pdv,Hu:2018veh,DiSalvo:2018ngq,Ray:2018hrx,Penalva:2019rgt,Mu:2019bin,Gutsche:2015mxa,Shivashankara:2015cta,Boer:2019zmp,Ferrillo:2019owd,Hu:2020axt}, $\Xi_b \to \Xi_c \tau \bar{\nu}_\tau$~\cite{Dutta:2018zqp,Faustov:2018ahb,Zhang:2019jax,Zhang:2019xdm}, $\Sigma_b\to \Sigma_c \tau \bar{\nu}_\tau$~\cite{Rajeev:2019ktp,Sheng:2020drc}, and $\Omega_b\to \Omega_c \tau \bar{\nu}_\tau$~\cite{Rajeev:2019ktp,Sheng:2020drc}.

Particularly, the LHCb collaboration released a value of the ratio $ R(J/\psi)\equiv\mathcal{B}(B_c\to J/\psi \tau \nu)/\mathcal{B}(B_c\to J/\psi \mu \nu)=0.71\pm 0.17\pm0.18$~\cite{Aaij:2017tyk}. Using the model-dependent calculations of $B_c \to J/\psi$ transition form factors~\cite{Kiselev:1999sc,Ivanov:2000aj,Ebert:2003cn,Hernandez:2006gt,Ivanov:2006ni,Wang:2008xt,Qiao:2012vt,Wen-Fei:2013uea,Rui:2016opu,Dutta:2017xmj,Watanabe:2017mip,Tran:2018kuv,Issadykov:2018myx,Leljak:2019eyw,Hu:2019qcn}, the SM prediction of $R(J/\psi)$ lies in the range 0.23--0.30. The model-independent bound on $R(J/\psi)$~\cite{Cohen:2018dgz,Wang:2018duy} is also obtained by constraining the $B_c \to J/\psi$ form factors through a combination of dispersive relations, heavy-quark relations at zero-recoil, and the limited existing form-factor determinations from lattice QCD~\cite{Colquhoun:2016osw,Lytle:2016ixw}, resulting in $0.20\leq R(J/\psi)\leq 0.39$~\cite{Cohen:2018dgz}. Very recently, the HPQCD collaboration present the first lattice QCD determination of the vector and axial-vector form factors of the $B_c \to J/\psi$ transition for the full $q^2$ range~\cite{Harrison:2020gvo}, and find $ R(J/\psi)=0.2582\pm0.0038$~\cite{Harrison:2020nrv} within the SM, which is the most accurate prediction and in tension with the LHCb result at $1.8\sigma$ level. The $B_c \to J/\psi$ transition form factors from lattice QCD make the theoretical calculations more accurate, which makes it interesting to revisit the NP effects in $B_c \to J/\psi \tau \bar{\nu}_\tau$ decay. 

In order to distinguish between the SM and NP scenarios, and further characterise the underlying effects of NP, besides considering the total decay rate, the full angular distribution of $B_c^- \to J/\psi \tau^-\bar{\nu}_\tau$ and $B_c^- \to J/\psi (\to \mu^+ \mu^-)\tau^-\bar{\nu}_\tau$ decays should also be taken into account sometimes, see for instance  refs.~\cite{Cohen:2018vhw,Leljak:2019eyw,Harrison:2020nrv}. However, as pointed out in refs.~\cite{Nierste:2008qe,Tanaka:2010se,Hagiwara:2014tsa,Bordone:2016tex,Alonso:2016gym,Alonso:2017ktd,Asadi:2018sym,Alonso:2018vwa,Bhattacharya:2020lfm,Asadi:2020fdo,Hu:2020axt,Ligeti:2016npd}, the information of the polar and azimuthal angles of the emitted $\tau^-$ cannot be determined precisely because the decay products of $\tau^-$ inevitably contain an undetected $\nu_\tau$. This means that the observables depending on the polar or azimuthal angle of $\tau^-$, such as the corresponding coefficients of the angular distribution and the forward-backward asymmetry of $\tau^-$, cannot be directly measured. Therefore, in this work, we construct a {\it measurable} angular distribution by further considering the subsequent decay $\tau^- \to \pi^- \nu_\tau$. The full cascade decay is $B_c^- \to J/\psi (\to \mu^+ \mu^-)\tau^- (\to \pi^- \nu_\tau)\bar{\nu}_\tau$, which includes three visible final-state particles $\mu^+$, $\mu^-$, and $\pi^-$ whose  three-momenta can be measured. We first calculate the full five-fold differential angular distribution including all Lorentz structures of the NP effective operators, and then carefully study the NP effects in the angular distribution from many aspects.

Our paper is organized as follows. In section~\ref{sec:analytical}, after defining the effective Hamiltonian, we give the analytical results of the independent transversity amplitudes and the measurable angular distribution of the five-body $B_c^- \to J/\psi (\to \mu^+ \mu^-)\tau^- (\to \pi^- \nu_\tau)\bar{\nu}_\tau$ decay. Definitions of the integrated observables are included in section~\ref{sec:obs}. In section~\ref{sec:Num}, we show the numerical results of the entire set of normalized angular observables $\widehat{\mathcal{I}}_i(q^2)$ and the lepton-flavor-universality ratios $R(P_{L,T}^{J/\psi})(q^2)$ and $R(J/\psi)$. A model-independent method for determining the existence of tensor operator is given in section~\ref{sec:sym}. Our conclusions are finally made in section~\ref{sec:conclusions}. In the appendices~\ref{sec:ang_dis} and \ref{sec:dep}, we present the detailed procedures related to the calculations of angular distribution and dependence relations, respectively.

\section{Analytical results}
\label{sec:analytical}

In this section, after giving some necessary definitions, we directly list the analytical results of angular distribution. The more detailed calculations, including some useful conventions, can be found in appendix~\ref{sec:ang_dis}.

\subsection{Effective Hamiltonian}
\label{subsec:EL}

Assuming that the NP scale is higher than the electroweak scale, one can integrate out the possible NP particles as well as the SM heavy particles --- the $W^\pm$, $Z^0$, the top quark, and the Higgs boson, thus obtaining the effective Hamiltonian suitable for describing the $b \to c \tau^- \bar{\nu}_\tau$ transition\footnote{Neutrinos are assumed to be left-handed in this work. The effective Hamiltonian containing right-handed neutrinos can be found in refs.~\cite{Dutta:2013qaa,Ligeti:2016npd,Mandal:2020htr}. It can be derived from the identity $\sigma^{\mu\nu}\gamma_5 = -\frac{i}{2}\epsilon^{\mu\nu\alpha\beta} \sigma_{\alpha\beta}$ that the operator $({\bar c}\sigma^{\mu\nu}(1+\gamma_5) b)({\bar\tau}\sigma_{\mu\nu} \nu_{\tau L}) $ is absent. We use the convention $\epsilon_{0123} = - \epsilon^{0123} = 1$.}
\begin{align}
	\label{eq:Heff}
	{\cal H}_{\rm eff} 
	= \sqrt{2} G_F V_{cb} \big[  
	& g_V ({\bar c}\gamma^\mu b) ({\bar\tau}\gamma_\mu \nu_{\tau L}) + g_A ({\bar c}\gamma^\mu \gamma_5 b)({\bar\tau}\gamma_\mu \nu_{\tau L}) 
	\nonumber\\
	&+ g_S ({\bar c} b)({\bar\tau} \nu_{\tau L}) + g_P ({\bar c} \gamma_5 b)({\bar\tau} \nu_{\tau L}) 
	\nonumber\\
	&+ g_T ({\bar c}\sigma^{\mu\nu}(1-\gamma_5) b)({\bar\tau}\sigma_{\mu\nu} \nu_{\tau L}) \big] + {\rm H.c.},
\end{align}
where $G_F$ is the Fermi constant, $V_{cb}$ is the CKM matrix element, $\sigma^{\mu\nu} \equiv \frac{i}{2}[\gamma^\mu,\, \gamma^\nu]$, and $\nu_{\tau L} = P_L \nu_\tau$ denotes the field of left-handed neutrino. The NP effects are encoded in the Wilson coefficients $g_i$, which are defined at the typical energy scale $\mu = m_b$. In the SM, $g_V = - g_A = 1$ and $g_S = g_P = g_T = 0$.

\subsection{Transversity amplitudes}
\label{subsec:TA}
In the calculation, the hadronic matrix elements contain the nonperturbative QCD effects and can be parameterized as the Lorentz invariant form factors. The vector and axial-vector current matrix elements can be written as the following four form factors~\cite{Beneke:2000wa,Sakaki:2013bfa,Harrison:2020gvo}
\begin{align}
	\langle  J/\psi(k,\varepsilon)|\bar{c}\gamma_\mu  b|B_c(p)\rangle &=
	\frac{2i V(q^2)}{m_{B_c} + m_{J/\psi}} \epsilon_{\mu\nu\alpha\beta}\varepsilon^{*\nu} k^\alpha p^\beta,  \\
	\langle  J/\psi(k,\varepsilon)|\bar{c}\gamma_\mu \gamma_5 b|B_c(p)\rangle& = \nonumber
	2m_{J/\psi}A_0(q^2)\frac{\varepsilon^* \cdot q}{q^2} q_\mu \nonumber\\
	&  +(m_{B_c}+m_{J/\psi})A_1(q^2)\left( \varepsilon^*_\mu - \frac{\varepsilon^* \cdot q}{q^2} q_\mu \right)\nonumber \\
	&-A_2(q^2)\frac{\varepsilon^* \cdot q}{m_{B_c}+m_{J/\psi}}\left( p_\mu + k_\mu - \frac{m_{B_c}^2-m_{J/\psi}^2}{q^2}q_\mu \right),
	\label{formfactors}
\end{align}
where $q=p-k$, $\varepsilon^{\mu}$ denotes the polarization vector of $J/\psi$ meson. In our numerical analysis, we will use the vector and axial-vector form factors computed in lattice QCD~\cite{Harrison:2020gvo,Harrison:2020nrv}. 

Using the equation of motion, the scalar and pseudo-scalar matrix elements can be obtained by
\begin{align}
 \langle  J/\psi(k,\varepsilon)|\bar{c}  b|B_c(p)\rangle &=0,
  \\
 \langle  J/\psi(k,\varepsilon)|\bar{c} \gamma_5 b|B_c(p)\rangle&=-\varepsilon^* \cdot q\frac{2m_{J/\psi}}{m_b+m_c}A_0(q^2),
\end{align}

Based on the above four form factors $V(q^2)$ and $A_{0,1,2}(q^2)$, one can define four independent transversity amplitudes as follows
\begin{align}
	\mathcal{A}_t&=\mathcal{A}^{SP}_t+\frac{m_\tau}{\sqrt{q^2}}\mathcal{A}^{VA}_t, \\
	\mathcal{A}_0&=g_A\frac{m_{B_c}+ m_{J/\psi}}{2m_{J/\psi}\sqrt{q^2}}\left[A_1(q^2)(m_{B_c}^2- m_{J/\psi}^2-q^2)-A_2(q^2)\frac{Q_+Q_-}{(m_{B_c}+ m_{J/\psi})^2}\right], \\
	\mathcal{A}_{\perp}&=g_A\sqrt{2}A_1(q^2)(m_{B_c}+ m_{J/\psi}), \\
	\mathcal{A}_{\parallel}&=g_V\sqrt{2}V(q^2)\frac{\sqrt{Q_+Q_-}}{m_{B_c}+ m_{J/\psi}},
	\label{transvers:VASP}
\end{align}
with
\begin{align}
	\mathcal{A}^{SP}_t=-g_PA_0(q^2)\frac{\sqrt{Q_+Q_-}}{m_b+m_c},\
	\mathcal{A}^{VA}_t=g_AA_0(q^2)\frac{\sqrt{Q_+Q_-}}{\sqrt{q^2}},
\end{align}
where $m_b$ and $m_c$ are the current quark masses evaluated at the scale $\mu= m_b$, and  $Q_\pm \equiv (m_{B_c}\pm  m_{J/\psi})^2 - q^2$.

The tensor matrix element can be parameterized as~\cite{Beneke:2000wa,Sakaki:2013bfa,Leljak:2019eyw}
\begin{align}
	\langle J/\psi(k, \varepsilon)| \bar c  \sigma_{\mu\nu}  b |B_c(p)\rangle =& \epsilon_{\mu \nu \alpha\beta } 
	\Bigg\{-\varepsilon^{*\alpha}(k + p)^\beta T_1(q^2) \nonumber\\
	& + \frac{m_{B_c}^2-m_{J/\psi}^2}{q^2}\varepsilon^{*\alpha} q^\beta \left[T_1(q^2) - T_2(q^2)\right]\nonumber\\
	& +  2 \frac{\varepsilon^* \cdot q }{q^2 }   p^\alpha k^\beta 
	\left[T_1(q^2) -T_2(q^2) - \frac{q^2}{m_{B_c}^2-m_{J/\psi}^2} T_3(q^2)\right] \Bigg\},
	\label{tformfactors}
\end{align}
and $\langle J/\psi | \bar c  \sigma_{\mu\nu} \gamma_5 b |B_c \rangle = -\frac{i}{2} \epsilon_{\mu \nu \alpha \beta} \langle J/\psi | \bar c  \sigma^{\alpha \beta} b |B_c\rangle$. In the presence of the tensor operators, we find three additional independent transversity amplitudes as follows
\begin{align}
	\mathcal{A}^T_0&=g_T\frac{1}{2m_{J/\psi}}\left[T_2(q^2)(m_{B_c}^2+3 m_{J/\psi}^2-q^2)-T_3(q^2)\frac{Q_+Q_-}{ m_{B_c}^2-m_{J/\psi}^2}\right],
	\\
	\mathcal{A}^T_{\perp}&=g_T\sqrt{2}T_2(q^2)\frac{m_{B_c}^2- m_{J/\psi}^2}{\sqrt{q^2}},
	\\
    \mathcal{A}^T_{\parallel}&=g_T\sqrt{2}T_1(q^2)\frac{\sqrt{Q_+Q_-}}{\sqrt{q^2}},
\end{align}
where the superscript $T$ indicates that an amplitude appears only when one considers the tensor operators.

\begin{figure}[t]
	\centering
	\includegraphics[width=0.45\textwidth]{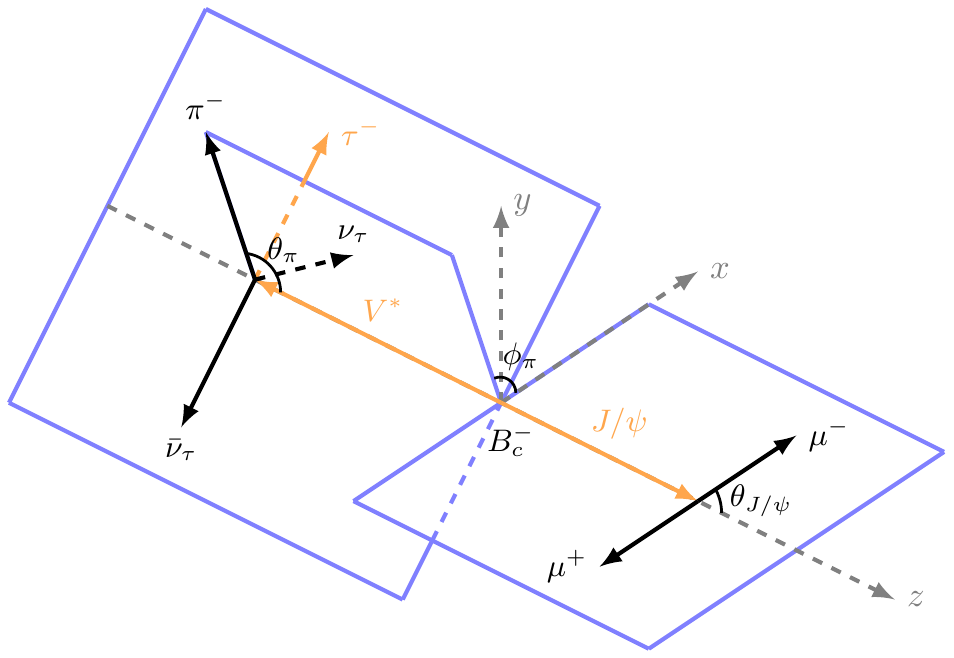}
	\caption{\label{fig:angle} \small Definition of the angles in the $B_c^- \to J/\psi (\to \mu^+ \mu^-)\tau^- (\to \pi^- \nu_\tau)\bar{\nu}_\tau$ decay.}
\end{figure}

\subsection{Angular distribution}
\label{sec:Angular distribution}

The {\it measurable} angular distribution of the five-body  $B_c^- \to J/\psi (\to \mu^+ \mu^-)\tau^- (\to \pi^- \nu_\tau)\bar{\nu}_\tau$ decay can be written as

\begin{align}
\label{eq:five-fold}
\frac{d^{5} \Gamma}{d q^{2} d E_{\pi} d \cos \theta_{\pi} d \phi_{\pi} d \cos \theta_{J/\psi}}=& \frac{3G_{F}^{2}\left|V_{c b}\right|^{2}\left|\boldsymbol{p}_{J/\psi}\right|\left(q^{2}\right)^{3 / 2} m_{\tau}^{2}}{256 \pi^{4} m_{B_c}^{2}\left(m_{\tau}^{2}-m_{\pi}^{2}\right)^{2}} {\cal B}_\tau {\cal B}_{J/\psi} \nonumber\\
& \times \mathcal{I}\left(q^{2}, E_{\pi}, \cos \theta_{J/\psi}, \cos \theta_{\pi}, \phi_{\pi}\right),
\end{align}
where $\left|\boldsymbol{p}_{J/\psi}\right|=\sqrt{Q_+Q_-}/(2 m_{B_c})$ denotes the magnitude of three-momentum of the $J/\psi$ meson in the $B_c$ rest frame. ${\cal B}_\tau \equiv {\cal B}\left(\tau^- \rightarrow \pi^{-} \nu_{\tau}\right)$ and ${\cal B}_{J/\psi} \equiv {\cal B}\left(J/\psi \rightarrow \mu^-\mu^+\right)$ are the branching fractions of $\tau^- \rightarrow \pi^{-} \nu_{\tau}$ and $J/\psi \rightarrow \mu^-\mu^+$ decays, respectively. Here $q^2$ is the invariant mass squared of the $\tau^-\bar{\nu}_\tau$ pair; $\theta_{J/\psi}$ denotes the polar angle of $\mu^-$ in the $J/\psi$ rest frame; $E_{\pi}$, $\theta_{\pi}$, and $\phi_{\pi}$ represent the energy, polar angle, and azimuthal
angle of $\pi^-$ in the $\tau^-\bar{\nu}_\tau$ center-of-mass frame, respectively. A more intuitive definition of the angles is shown in figure~\ref{fig:angle}. The function $\mathcal{I}\left(q^{2}, E_{\pi}, \cos \theta_{J/\psi}, \cos \theta_{\pi}, \phi_{\pi}\right)$ can be decomposed into a set of trigonometric functions as follows
\begin{align}
	\mathcal{I}\left(q^{2}, E_{\pi}, \cos \theta_{J/\psi}, \cos \theta_{\pi}, \phi_{\pi}\right)
	=&\sum_{i=1}^{12}\mathcal{I}_i\left(q^{2}, E_\pi \right)\Omega_i\left(\cos \theta_{J/\psi}, \cos \theta_{\pi}, \phi_{\pi}\right)\nonumber\\
	\equiv
	& \mathcal{I}_{1 c} \cos ^{2} \theta_{J/\psi}+\mathcal{I}_{1s} \sin ^{2} \theta_{J/\psi} \nonumber\\
	&+\left(\mathcal{I}_{2 c} \cos ^{2} \theta_{J/\psi}+\mathcal{I}_{2 s} \sin ^{2} \theta_{J/\psi}\right) \cos 2 \theta_{\pi} \nonumber\\
	&+\left(\mathcal{I}_{6 c} \cos ^{2} \theta_{J/\psi}+\mathcal{I}_{6 s} \sin ^{2} \theta_{J/\psi}\right) \cos \theta_{\pi} \nonumber\\
	&+\left(\mathcal{I}_{3} \cos 2 \phi_{\pi}+\mathcal{I}_{9} \sin 2 \phi_{\pi}\right) \sin ^{2} \theta_{\pi} \sin ^{2} \theta_{J/\psi} \nonumber\\
	&+\left(\mathcal{I}_{4} \cos \phi_{\pi}+\mathcal{I}_{8} \sin \phi_{\pi}\right) \sin 2 \theta_{\pi} \sin 2 \theta_{J/\psi} \nonumber\\
	&+\left(\mathcal{I}_{5} \cos \phi_{\pi}+\mathcal{I}_{7} \sin \phi_{\pi}\right) \sin \theta_{\pi} \sin 2 \theta_{J/\psi}.
\end{align}
The twelve angular observables $\mathcal{I}_{i} (q^2,E_{\pi})$ can be completely expressed in terms of the seven transversity amplitudes defined in subsection~\ref{subsec:TA} and the dimensionless factors listed in appendix~\ref{subsec:DF}. Explicitly, we have
\begin{align}
	\mathcal{I}_{1 c} =&S_1 \left| \mathcal{A}_{\perp}\right|
	^2-2 R_1 {\rm Re}\left[\mathcal{A}_{\perp} \mathcal{A}_{\perp}^{T*} \right]+4S_1^T \left| \mathcal{A}_{\perp}^{T}\right| ^2+(\perp \leftrightarrow \parallel),\\[2mm]
	\mathcal{I}_{1s}=&S_t\left| \mathcal{A}_{t}\right| ^2+(S_1-S_3)\left|\mathcal{A}_{0}\right|^2 -2 (R_1-R_3) {\rm Re}\left[\mathcal{A}_{0} \mathcal{A}_{0}^{T*} \right] +4(S^T_1-S^T_3) \left|\mathcal{A}_{0}^{T}\right|^2
	\nonumber\\
	&+\frac{1}{2}\left\{S_1\left| \mathcal{A}_{\perp}\right|^2- 2R_1 {\rm Re}\left[\mathcal{A}_{\perp} \mathcal{A}_{\perp}^{T*} \right]+4S_1^T \left| \mathcal{A}_{\perp}^{T}\right| ^2+(\perp \leftrightarrow \parallel)\right\},\\[2mm]
	\mathcal{I}_{2c}=&S_3 \left| \mathcal{A}_{\perp}\right|
	^2-2 R_3 {\rm Re}\left[\mathcal{A}_{\perp} \mathcal{A}_{\perp}^{T*}\right]+ 4S^T_3 \left|\mathcal{A}_{\perp}^{T}\right| ^2+ (\perp \leftrightarrow \parallel),\\[2mm]
	\mathcal{I}_{2s}=&-2S_3 \left| \mathcal{A}_{0}\right| ^2+4 R_3{\rm Re}\left[\mathcal{A}_{0}\mathcal{A}_{0}^{T*}\right] -8 S^T_3 \left|
	\mathcal{A}_{0}^{T}\right| ^2\nonumber\\
	&+\frac{1}{2}\left\{S_3 \left| \mathcal{A}_{\perp}\right| ^2- 2 R_3  {\rm Re}\left[\mathcal{A}_{\perp} \mathcal{A}_{\perp}^{T*}\right]+ 4S^T_3 \left|\mathcal{A}_{\perp}^{T}\right| ^2+ (\perp \leftrightarrow \parallel)\right\},\\[2mm]
	\mathcal{I}_{6c}=& 2\mathrm{Re}\left[S_2 \mathcal{A}_{\parallel}\mathcal{A}_{\perp}^* - R_2\left(\mathcal{A}_{\perp} \mathcal{A}_{\parallel}^{T*}+\mathcal{A}_{\parallel} \mathcal{A}_{\perp}^{T*}\right)+4S_2^T \mathcal{A}_{\parallel}^{T}
	\mathcal{A}_{\perp}^{T*}\right],\\[2mm]
	\mathcal{I}_{6s}=&\mathrm{Re}\Big[S_2\mathcal{A}_{\parallel}
	\mathcal{A}_{\perp}^*- R_2\left(\mathcal{A}_{\perp} \mathcal{A}_{\parallel}^{T*}+ \mathcal{A}_{\parallel}
	\mathcal{A}_{\perp}^{T*}\right)+4S_2^T\mathcal{A}_{\parallel}^{T}
	\mathcal{A}_{\perp}^{T*}\nonumber\\
	&-\sqrt{2}R_t\mathcal{A}_{t} \mathcal{A}_{0}^*+2\sqrt{2}R_t^T \mathcal{A}_{t}\mathcal{A}_{0}^{T*}\Big],\\[2mm]
	\mathcal{I}_{3}=&S_3 \left|
	\mathcal{A}_{\perp}\right| ^2-2R_3\mathrm{Re}\left[\mathcal{A}_{\perp} \mathcal{A}_{\perp}^{T*}\right]+4S^T_3 \left| \mathcal{A}_{\perp}^{T}\right| ^2-(\perp \leftrightarrow \parallel),\\[2mm]
	\mathcal{I}_{9}=& 2\mathrm{Im}\left[S_3\mathcal{A}_{\parallel} \mathcal{A}_{\perp}^* + R_3\left(\mathcal{A}_{\perp} \mathcal{A}_{\parallel}^{T*}-\mathcal{A}_{\parallel} \mathcal{A}_{\perp}^{T*}\right) +4S^T_3 \mathcal{A}_{\parallel}^{T}\mathcal{A}_{\perp}^{T*}\right],\\[2mm]
	\mathcal{I}_{4}=&\sqrt{2}\mathrm{Re}\left[S_3\mathcal{A}_{\perp} \mathcal{A}_{0}^* - R_3\left(\mathcal{A}_{0} \mathcal{A}_{\perp}^{T*}+ \mathcal{A}_{\perp} \mathcal{A}_{0}^{T*}\right) +4S^T_3
	\mathcal{A}_{0}^{T} \mathcal{A}_{\perp}^{T*}\right],\\[2mm]
	\mathcal{I}_{8}=& \sqrt{2}\mathrm{Im}\left[S_3\mathcal{A}_{\parallel} \mathcal{A}_{0}^* + R_3\left(\mathcal{A}_{0} \mathcal{A}_{\parallel}^{T*} - \mathcal{A}_{\parallel}\mathcal{A}_{0}^{T*}\right) + 4 S^T_3
	\mathcal{A}_{\parallel}^{T} \mathcal{A}_{0}^{T*} \right],\\[2mm]
	\mathcal{I}_{5}=& \frac{1}{2\sqrt{2}}\mathrm{Re}\Big[2S_2\mathcal{A}_{\parallel}\mathcal{A}_{0}^* - 2R_2 \left(\mathcal{A}_{0}
	\mathcal{A}_{\parallel}^{T*}+\mathcal{A}_{\parallel}\mathcal{A}_{0}^{T*}\right) + 8S_2^T \mathcal{A}_{\parallel}^{T}\mathcal{A}_{0}^{T*}\nonumber\\
	&+ \sqrt{2}R_t\mathcal{A}_{t}\mathcal{A}_{\perp}^* - 2\sqrt{2}R_t^T\mathcal{A}_{t}\mathcal{A}_{\perp}^{T*} \Big],\\[2mm]
	\mathcal{I}_{7}=&\frac{1}{2\sqrt{2}}\mathrm{Im}\Big[2S_2\mathcal{A}_{\perp}\mathcal{A}_{0}^* + 2R_2\left(\mathcal{A}_{0} \mathcal{A}_{\perp}^{T*} -\mathcal{A}_{\perp}
	\mathcal{A}_{0}^{T*}\right) - 8S_2^T \mathcal{A}_{0}^{T} \mathcal{A}_{\perp}^{T*} \nonumber\\
	&-\sqrt{2}R_t\mathcal{A}_{t}\mathcal{A}_{\parallel}^* + 2\sqrt{2}R_t^T\mathcal{A}_{t}\mathcal{A}_{\parallel}^{T*}\Big].
\end{align}

In the SM, the angular observables $\mathcal{I}_{7}$, $\mathcal{I}_{8}$, and $\mathcal{I}_{9}$ are vanishing. Therefore, in future measurements, a non-vanishing $\mathcal{I}_{7}$, $\mathcal{I}_{8}$, or $\mathcal{I}_{9}$ would be a solid signal of NP, which induces a complex contribution to the amplitude.

\section{Integrated observables}
\label{sec:obs}

\subsection{$E_\pi$-integrated angular observables}
The differential decay rate \eqref{eq:five-fold} depends on five parameters $q^2$, $E_\pi$, $\theta_{J/\psi}$, $\theta_{\pi}$ and $\phi_\pi$, and a complete experimental analysis may be limited by statistics. Integrating over the $E_\pi$ and after a proper normalization, we can get the following angular function
\begin{align}\label{eq:afq2}
	\widehat{\mathcal{I}}\left(q^{2}, \cos \theta_{J/\psi}, \cos \theta_{\pi}, \phi_{\pi}\right)\equiv&\frac{\int{\frac{d^{5} \Gamma}{d q^{2} d E_{\pi} d \cos \theta_{\pi} d \phi_{\pi} d \cos \theta_{J/\psi}}}dE_\pi}{\frac{d\Gamma}{dq^2}}\nonumber\\
	=&\frac{9}{8\pi}\sum_{i=1}^{12}\widehat{\mathcal{I}}_i\left(q^{2}\right)\Omega_i\left(\cos \theta_{J/\psi}, \cos \theta_{\pi}, \phi_{\pi}\right),
\end{align}
with the twelve normalized angular observables $\widehat{\mathcal{I}_i}(q^2)$ defined as
\begin{align}
	\widehat{\mathcal{I}}_i\left(q^{2}\right)\equiv\frac{\int{\mathcal{I}_i\left(q^{2},E_\pi\right)}dE_\pi}{\int{\left(3\mathcal{I}_{1c}-\mathcal{I}_{2c}+6\mathcal{I}_{1s}-2\mathcal{I}_{2s}\right)}dE_\pi}.
\end{align}
Our choice of the normalization in eq.~\eqref{eq:afq2} results the relationship $3\widehat{\mathcal{I}}_{1c}\left(q^2\right)-\widehat{\mathcal{I}}_{2c}\left(q^2\right)+6\widehat{\mathcal{I}}_{1s}\left(q^2\right)-2\widehat{\mathcal{I}}_{2s}\left(q^2\right)=1$. The cancellations through normalization to the decay rate lead to the observation that the observables $\widehat{\mathcal{I}_i}(q^2)$ have less theoretical uncertainty to facilitate the discussion of the NP effects. In section~\ref{sec:Num}, we will analyze numerically the entire set of observables $\widehat{\mathcal{I}_i}(q^2)$ within the SM and in some NP benchmark points.

The forward-backward asymmetry of $\pi^-$ meson as a function of $q^2$ can be defined as 
\begin{align}
	A_{FB} (q^2) 
	\equiv & 
	\frac{\int_{0}^{1} \frac{d^2\Gamma}{dq^2 d\cos\theta_\pi} d\cos\theta_\pi - \int_{-1}^{0} \frac{d^2\Gamma}{dq^2 d\cos\theta_\pi} d\cos\theta_\pi}{\frac{d\Gamma}{dq^2}}\nonumber\\
	=&
	\frac{3}{2}\left(\widehat{\mathcal{I}}_{6c} + 2\widehat{\mathcal{I}}_{6s}\right).
	\label{eq:AFB}
\end{align}
This asymmetry observable only exists in $\tau$ channel, and specifically for the $\tau^- \to \pi^- \nu_\tau$ decay. Obviously, it can be expressed linearly in terms of angular observables $\widehat{\mathcal{I}_i}(q^2)$.

By integrating over the lepton-side parameters $E_\pi$, $\theta_{\pi}$, $\phi_{\pi}$, one can obtain the two-fold differential decay rate as follows
\begin{align}
	\frac{d^{2} \Gamma}{d q^{2} d\cos \theta_{J/\psi}}=\frac{3}{8}\frac{d \Gamma}{d q^{2} }\left[2P_{L}^{J/\psi}\left(q^2\right)\sin^2\theta_{J/\psi}+P_{T}^{J/\psi}\left(q^2\right)\left(1+\cos^2\theta_{J/\psi}\right)\right],
\end{align}
where 
\begin{align}
	P_{L}^{J/\psi}\left(q^2\right)\equiv\frac{d\Gamma_L / dq^2}{d\Gamma_L / dq^2+d\Gamma_T / dq^2},\quad P_{T}^{J/\psi}\left(q^2\right)\equiv\frac{d\Gamma_T / dq^2}{d\Gamma_L / dq^2+d\Gamma_T / dq^2},
\end{align}
are the longitudinal and transverse polarization fractions of the $J/\psi$ meson, respectively. The differential decay rates for the longitudinally and transversely polarized intermediate state $J/\psi$ are given, respectively, by
\begin{align}
	\frac{d\Gamma_L}{dq^2}\equiv&
	\frac{d\Gamma^{\lambda_{J/\psi}=0}}{dq^2}\nonumber\\
	=&\mathcal{N}\bigg\{3 q^2 \left|
	\mathcal{A}_t\right| ^2+\left(m_\tau^2+2 q^2\right) \left| \mathcal{A}_0\right| ^2 \nonumber\\
	&+24 m_\tau
	\sqrt{q^2} \mathrm{Re}\left[\mathcal{A}_0 \mathcal{A}_0^{T*}\right]+16 \left(2m_\tau^2+q^2\right)
	\left| \mathcal{A}_0^{T}\right| ^2\bigg\},\label{eq:dGL}\\
	\frac{d\Gamma_T}{dq^2}\equiv&
	\frac{d\Gamma^{\lambda_{J/\psi}=+1}}{dq^2}+\frac{d\Gamma^{\lambda_{J/\psi}=-1}}{dq^2}\nonumber\\
	=&\mathcal{N}\bigg\{\left(m_\tau^2+2 q^2\right)\left| \mathcal{A}_\perp\right| ^2 
	+24 m_\tau
	\sqrt{q^2} \mathrm{Re}\left[\mathcal{A}_\perp \mathcal{A}_\perp^{T*}\right]\nonumber\\
	&+16\left(2 m_\tau^2+q^2\right) \left|\mathcal{A}_\perp^{T}\right| ^2+(\perp \leftrightarrow \parallel)\bigg\},\label{eq:dGT}
\end{align}
with the factor
\begin{align}
	\mathcal{N}\equiv\frac{G_{F}^{2}\left|V_{c b}\right|^{2}\left|\boldsymbol{p}_{J/\psi}\right|}{192 \pi^{3} m_{B_c}^{2}}\left(1-\frac{m_\tau^2}{q^2}\right)^2 \mathcal{B}_\tau \mathcal{B}_{J/\psi}.
\end{align}
The polarization observables $P_{L,T}^{J/\psi}(q^2)$ constructed above are not affected by $\tau$ decay dynamics since we have integrated over all the lepton-side kinematic parameters, so they are also applicable to light leptons $\mu$ and $e$. We denote $\left[P_{L,T}^{J/\psi}\right]_\tau$ and $\left[P_{L,T}^{J/\psi}\right]_\mu$ as extraction from $B_c\to J/\psi \tau \nu$ and $B_c\to J/\psi \mu \nu$ decays respectively, and define the following ratios to probe the universality of lepton flavor
\begin{equation}
R\left(P_{L,T}^{J/\psi} \right) \equiv \frac{\left[P_{L,T}^{J/\psi}\right]_\tau}{\left[P_{L,T}^{J/\psi}\right]_\mu}.
\end{equation}
The $q^2$ distribution of the decay rate can be obtained by adding up eqs.~\eqref{eq:dGL} and \eqref{eq:dGT} as follows
\begin{align}
	\frac{d\Gamma}{dq^2}=\frac{d\Gamma_L}{dq^2}+\frac{d\Gamma_T}{dq^2}.
\end{align}
Our $d\Gamma/dq^2$ (apart from ${\cal B}_\tau \mathcal{B}_{J/\psi}$) is consistent with that in refs.~\cite{Harrison:2020nrv,Sakaki:2013bfa}.

\subsection{Tau asymmetries in $B_c^- \to J/\psi \tau^- \bar{\nu}_\tau$ from the visible kinematics}

After integrating over the variables $\theta_{J/\psi}$ and $\phi_{\pi}$, one has
\begin{align}
	\label{eq:three-fold}
	\frac{d^{3} \Gamma}{d q^{2} d E_{\pi} d \cos \theta_{\pi}}=& \frac{G_{F}^{2}\left|V_{c b}\right|^{2}\left|\boldsymbol{p}_{J/\psi}\right|\left(q^{2}\right)^{3 / 2} m_{\tau}^{2}}{64 \pi^{3} m_{B_c}^{2}\left(m_{\tau}^{2}-m_{\pi}^{2}\right)^{2}} {\cal B}_\tau {\cal B}_{J/\psi} \nonumber\\
	& \times \left[\mathcal{I}_{1c} + 2\mathcal{I}_{1s}  + \left(\mathcal{I}_{6c} + 2\mathcal{I}_{6s} \right)\cos\theta_{\pi} + \left(\mathcal{I}_{2c} + 2\mathcal{I}_{2s}\right)\cos2\theta_{\pi}  \right] .
\end{align}
This three-fold differential decay rate can be used to indirectly reveal the information of $\tau$ asymmetries in $B_c^- \to J/\psi \tau^- \bar{\nu}_\tau$ decay~\cite{Kiers:1997zt,Nierste:2008qe,Tanaka:2010se,Sakaki:2012ft,Ivanov:2017mrj,Alonso:2017ktd,Asadi:2020fdo}. The following discussion of this subsection will follow closely refs.~\cite{Alonso:2017ktd,Asadi:2020fdo}, which give a detailed analysis of the $\tau$ properties in $B\to D^{(*)} \tau^- \bar{\nu}_\tau$ decays using $\tau\to d \nu (d=\pi, \, \rho)$.

The eq.~\eqref{eq:three-fold} can be rewritten as
\begin{equation}
\frac{d^{3} \Gamma}{d q^{2} d \omega_{\pi} d \cos \theta_{\pi}}= 
\frac{d\Gamma}{d q^{2}} \sum_{n=0}^{2} P_n\left(\cos\theta_{\pi} \right) I_n(q^2, \omega_{\pi}),
\end{equation}
with
\begin{align}
I_0 \equiv & \frac{1}{2}\left[f_0\left(q^2 \right) + f_L\left(q^2,\omega_{\pi} \right) P_L(q^2) \right],
\\
I_1 \equiv & f_{A_\tau} \left(q^2,\omega_{\pi} \right) A_\tau\left(q^2 \right) + f_\perp \left(q^2,\omega_{\pi} \right) P_\perp \left(q^2 \right) + f_{Z_L}\left(q^2,\omega_{\pi} \right) Z_L\left(q^2\right),
\\
I_2 \equiv & f_{Z_\perp}\left(q^2,\omega_{\pi} \right) Z_\perp\left(q^2\right) + f_{Z_Q}\left(q^2,\omega_{\pi} \right) Z_Q\left(q^2\right) + f_{A_Q}\left(q^2,\omega_{\pi} \right) A_Q\left(q^2\right).
\end{align}
Here, $P_{0,1,2}\left(\cos\theta_{\pi} \right)$ are the Legendre functions, the $\omega_{\pi}$ and the following $\kappa_{\pi}$ and $\kappa_{\tau}$ are defined in eq.~\eqref{eq:dimenless_paras}. The $\tau$ asymmetries $P_L$, $A_\tau$, $P_\perp$, $Z_L$, $Z_\perp$, $Z_Q$, and $A_Q$ are defined in the section 2 of ref.~\cite{Asadi:2020fdo}. We find that the functions $f_i$ are given, respectively, by
\begin{align}
f_0 = & g_+ + g_-, 
& 
f_L = & g_+ - g_-, 
& 
f_\perp =& \frac{4}{\pi} \sin^2\theta_{\pi\tau}\, h_\perp,\nonumber
\\
f_{A_\tau} =& \cos\theta_{\pi\tau}\, f_0,
&
f_{Z_L} = & \cos\theta_{\pi\tau}\, f_L,
&
f_{Z_\perp} = & \frac{3\pi}{4} \cos\theta_{\pi\tau}\, f_\perp,\nonumber
\\
f_{A_Q} = & \frac{1}{2}\left(3\cos^2\theta_{\pi\tau} -1\right) f_0,
&
f_{Z_Q} = & \frac{1}{2}\left(3\cos^2\theta_{\pi\tau} -1\right) f_L,
\end{align}
where $\cos\theta_{\pi\tau} = \frac{\kappa _{\tau }^2+\kappa _{\pi}^2-\omega _{\pi } \left(\kappa _{\tau }^2+1\right)}{\left(\kappa _{\tau }^2-1\right) \sqrt{\omega _{\pi }^2-\kappa _{\pi }^2}}$ is the cosine of the $\pi$-$\tau$ opening angle $\theta_{\pi\tau}$ in the $\tau^-\bar{\nu}_\tau$ center-of-mass frame~\cite{Hu:2020axt}, and 
\begin{align}
g_+ =& \frac{2 \kappa _{\tau }^2 \left(2 \omega _{\pi } \kappa _{\tau }^2-\kappa _{\tau }^4-\kappa
   _{\pi }^2\right)}{\left(\kappa _{\pi }^2-\kappa _{\tau }^2\right)^2 \left(\kappa _{\tau
   }^2-1\right)^2},
   \\
g_- = & \frac{2 \kappa _{\tau }^4 \left(\kappa _{\pi }^2-2 \omega _{\pi }+1\right)}{\left(\kappa _{\pi
   }^2-\kappa _{\tau }^2\right)^2 \left(\kappa _{\tau }^2-1\right)^2},
   \\
h_\perp = &  \frac{2 \kappa _{\tau }^3 \sqrt{\omega _{\pi }^2-\kappa _{\pi }^2}}{\left(\kappa _{\pi
   }^2-\kappa _{\tau }^2\right)^2 \left(\kappa _{\tau }^2-1\right)}.
\end{align}
Neglecting the $\pi$ mass, our results are in agreement with those in refs.~\cite{Alonso:2017ktd,Asadi:2020fdo}. The sign difference in $h_\perp$ is due to the different choice of reference direction. It should be pointed out that in the absence of $Z_L(q^2)$, the differential forward-backward asymmetry $dA_\pi/d\omega_\pi$ (i.e. $I_1(q^2,\omega_\pi)$) cannot be expressed in terms of $A_\tau(q^2)$ and $P_\perp(q^2)$ as given by eq.~(16) of ref.~\cite{Alonso:2017ktd}.

\section{Numerical results}
\label{sec:Num}

\subsection{The form factors}
\label{subsec:FFs}

The $B_c\to J/\psi$ transition form factors are the main source of theoretical uncertainties. For the $B_c\to J/\psi$ vector and axial-vector form factors, $V(q^2)$ and $A_{0,1,2}(q^2)$, we use the latest high-precision lattice QCD calculation results given in ref.~\cite{Harrison:2020gvo}. Since the $B_c\to J/\psi$ tensor form factors $T_{1,2,3}(q^2)$ are not included in ref.~\cite{Harrison:2020gvo}, we will adopt the $T_{1,2,3}(q^2)$ calculated in the QCD sum rule method~\cite{Leljak:2019eyw}.\footnote{We are very grateful to Domagoj Leljak for providing us with the variances and correlation matrix of $z$-expansion parameters of $B_c\to J/\psi$ tensor form factors.} These form factors are parameterized in a simplified $z$ expansion to extend to the full $q^2$ range.

\subsection{The NP benchmark points}
\label{subsec:NPBPs}

The model-independent analyses of NP effects in $B\to D^{(*)}\tau\nu$ decays have been completed in many previous works~\cite{Alok:2017qsi,Hu:2018veh,Alok:2019uqc,Murgui:2019czp,Blanke:2018yud,Blanke:2019qrx,Shi:2019gxi,Cheung:2020sbq,Kumbhakar:2020jdz}. In order to show the influences of these NP effects on the angular distribution of $B_c^- \to J/\psi (\to \mu^+ \mu^-)\tau^- (\to \pi^- \nu_\tau)\bar{\nu}_\tau$ decay, we select various best-fit values as the NP benchmark points. These best-fit values are usually performed on a set of chiral base, which is equivalent to Eq.~\eqref{eq:Heff} by the following relations
\begin{align}
	g_V =& 1+C_{V_L}+C_{V_R}, & g_A =& -1-C_{V_L}+C_{V_R}, & \nonumber\\
	g_S =& C_{S_L} + C_{S_R}, & g_P =& -C_{S_L} + C_{S_R}, & g_T = C_T.
\end{align}
According to the following steps, we select a total of eight NP benchmark points under seven different NP hypotheses.

Switching one coupling $C_i$ at a time, there are five NP hypotheses. The hypothesis of a single $C_{V_L}$ can resolve the $R(D^{(*)})$ anomalies well, but there is no effect on the normalized observables defined in section~\ref{sec:obs}, so we should not choose it. The hypothesis of a single $C_{S_L}$ or $C_{S_R}$ is ruled out by the decay rate of $B_c \to \tau \nu$ decay~\cite{Li:2016vvp,Celis:2016azn,Alonso:2016oyd}. We take a benchmark point from each of the two remaining NP hypotheses as follows~\cite{Cheung:2020sbq}
\begin{align*}
&\text{BP1:} \qquad \left({\rm Re}\left[C_{V_R}\right], \, {\rm Im}\left[C_{V_R}\right]\right) = \left(-0.030,\,  0.460\right)
\\
&\text{BP2:} \qquad  \left({\rm Re}\left[C_{T}\right], \, {\rm Im}\left[C_{T}\right]\right) = \left(0.011,\,  0.165\right)
\end{align*}
The corresponding complex-conjugated fitting values $\left({\rm Re}\left[C_{V_R}\right], \, {\rm Im}\left[C_{V_R}\right]\right) = \left(-0.030,\, -0.460\right)$ and $\left({\rm Re}\left[C_{T}\right], \, {\rm Im}\left[C_{T}\right]\right) = \left(0.011,\, -0.165\right)$ are marked as BP1$^*$ and BP2$^*$, respectively. Although BP1~(BP2) and BP1$^*$~(BP2$^*$) are formally different benchmark points, they produce the same results for angular observables $\widehat{{\cal I}}_{1c,1s,2c,2s,6c,6s,3,4,5}$ and opposite results for $\widehat{{\cal I}}_{7,8,9}$. Observables $\widehat{{\cal I}}_{7,8,9}$ can distinguish between the NP benchmark point and its complex conjugate partner very well. In the following analysis, we do not consider BP1$^*$ and BP2$^*$, and the same treatment is also applicable to the following BP6$^*$, which is the complex conjugate of the benchmark point BP6.

Considering the combinations induced by specific UV models, we choose the  best-fit points in the following four different NP hypotheses as our NP benchmark points (the remaining $C_i$ are set to zero in each case)~\cite{Blanke:2019qrx}
\begin{align*}
	\text{BP3:} \qquad &  \left(C_{V_L},\, C_{S_L} = -4 C_T\right)= \left(0.10, \, -0.04\right)
	\\
	\text{BP4:} \qquad  &  \left(C_{S_R},\, C_{S_L}\right)= \left(0.21, \, -0.15\right)\, \text{(A)} \quad \text{or}  \quad  \left(-0.26, \, -0.61\right)\, \text{(B)}
	\\
	\text{BP5:} \qquad &  \left(C_{V_L},\, C_{S_R} \right)= \left(0.08, \, -0.01\right)
	\\
	\text{BP6:} \qquad &  \left({\rm Re}\left[C_{S_L} = 4 C_T\right],\, {\rm Im}\left[C_{S_L} = 4 C_T\right] \right)= \left(-0.06, \, 0.31\right)
\end{align*}
where the Wilson coefficients are given at the NP scale 1TeV, and we should run them down to the scale $m_b$~\cite{Blanke:2018yud}.

Finally, taking into account all NP Wilson coefficients, except $C_{V_R}$ which is explicitly lepton-flavor universal in the standard model effective field theory formalism up to contributions of ${\cal O}(\mu^4_{\rm EW}/\Lambda^4)$~\cite{Hu:2018veh}, we choose a set of values labelled ``Min 1b" in table 8 of ref.~\cite{Murgui:2019czp} as our NP benchmark point BP7
\begin{equation*}
	\text{BP7:} \qquad   \left(C_{V_L},\, C_{S_R}, \, C_{S_L}, \, C_T \right)= \left(0.09, \, 0.086,\, -0.14,\, 0.008\right)
\end{equation*}
We adopt the same treatment as in many literatures (e.g.~\cite{Becirevic:2019tpx,Boer:2019zmp,Asadi:2020fdo,Harrison:2020nrv,Alguero:2020ukk}), that is, only the central value of best-fit result is considered as the benchmark point to qualitatively discuss the influence of the NP effect. 

\subsection{Angular observables $\widehat{\cal I}_i(q^2)$}

%%%%%%%%%%%%%%%%%%%%%%%%%%%%%%%%%
\begin{figure}[t]
\centering
\includegraphics[width=0.31\textwidth]{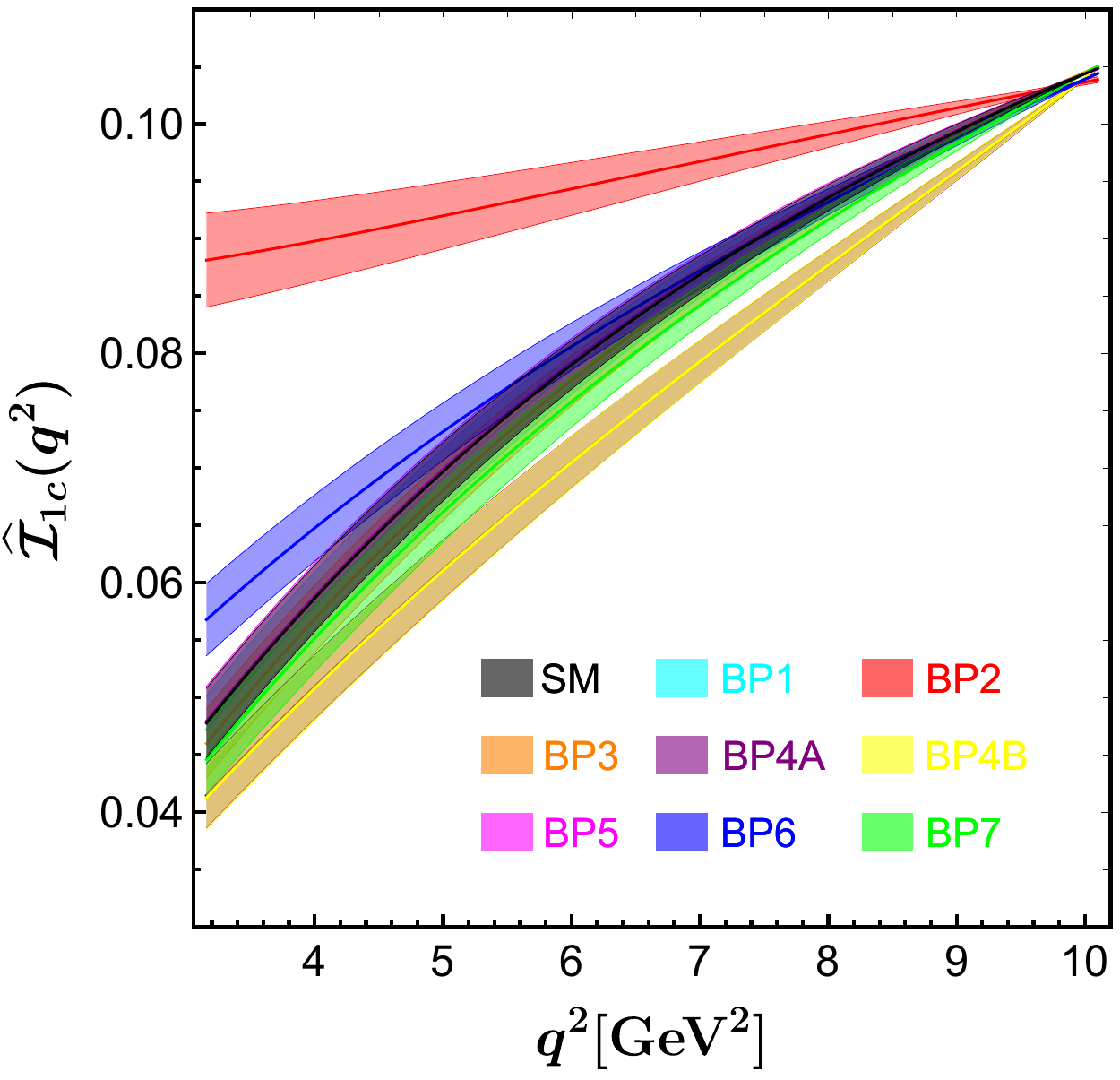}
\includegraphics[width=0.31\textwidth]{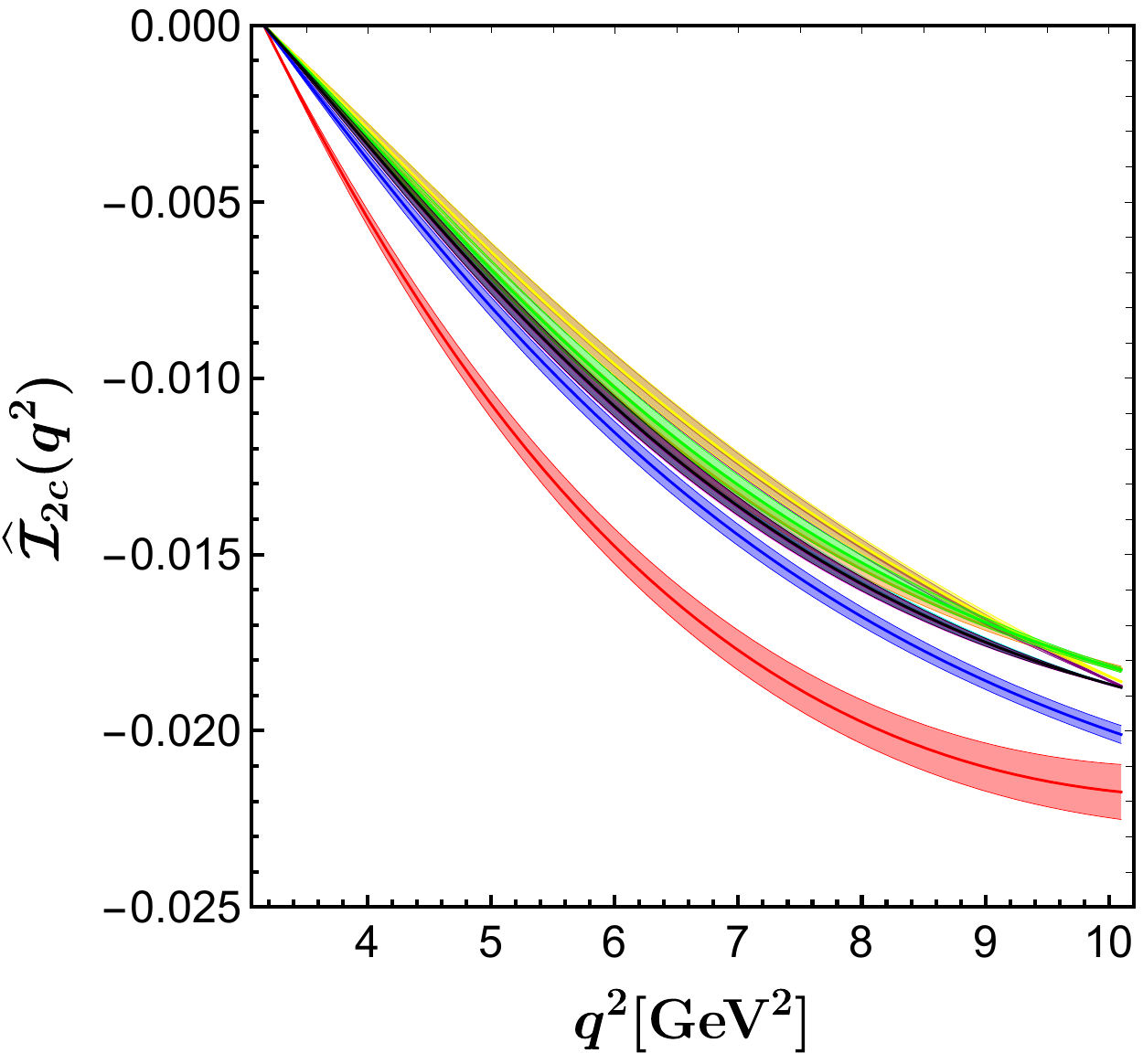}
\includegraphics[width=0.31\textwidth]{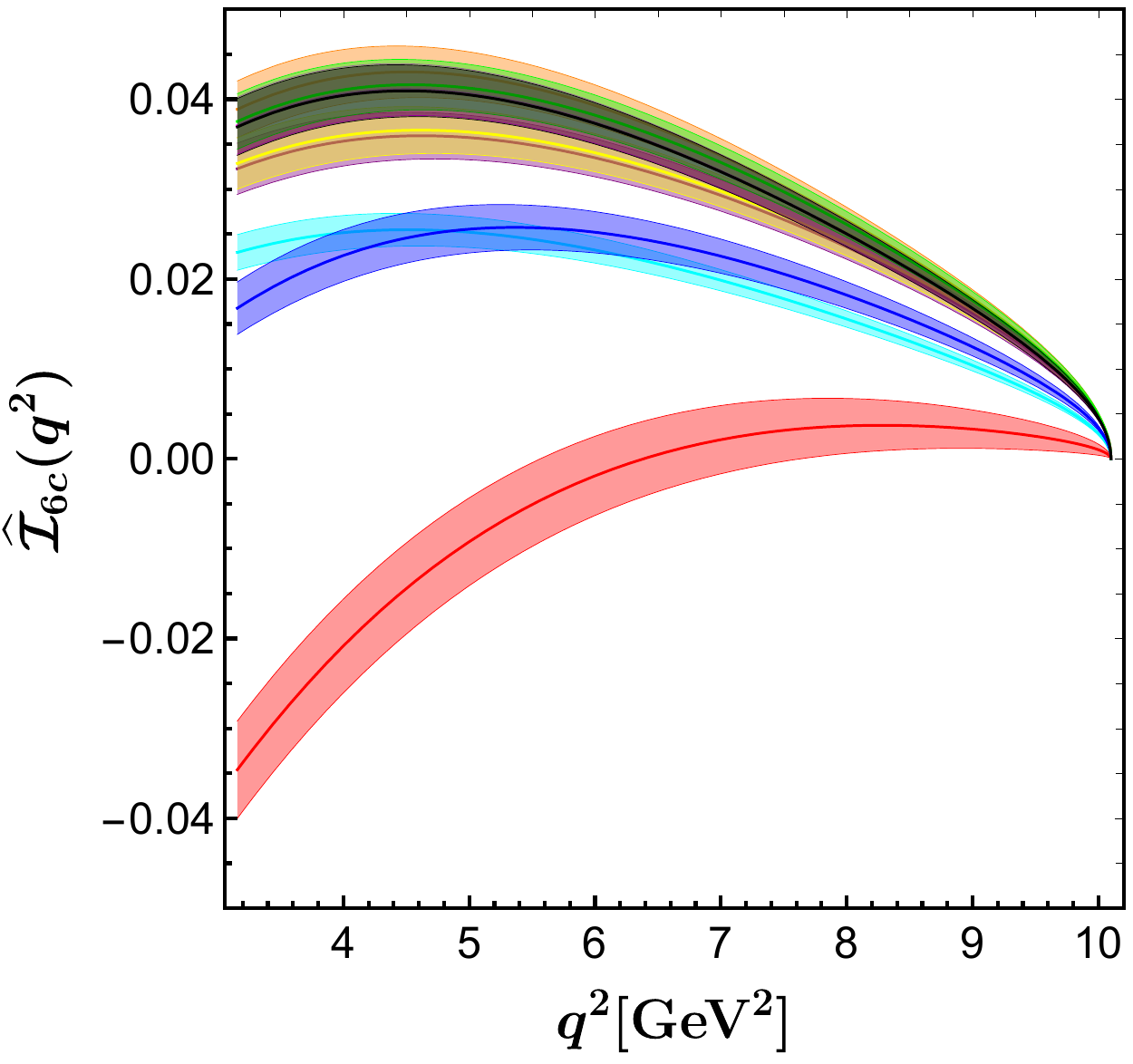}
\\
\includegraphics[width=0.31\textwidth]{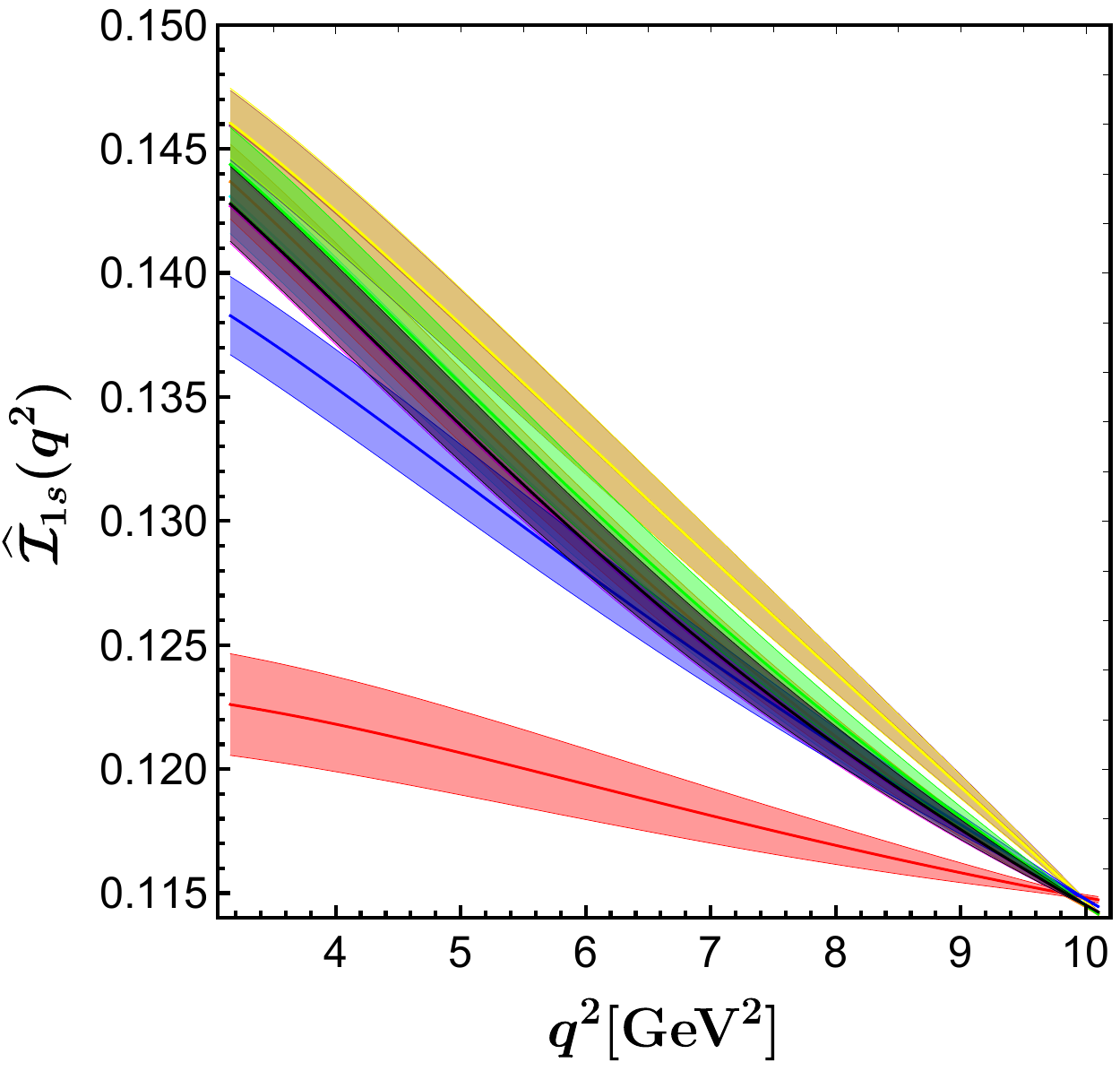}
\includegraphics[width=0.31\textwidth]{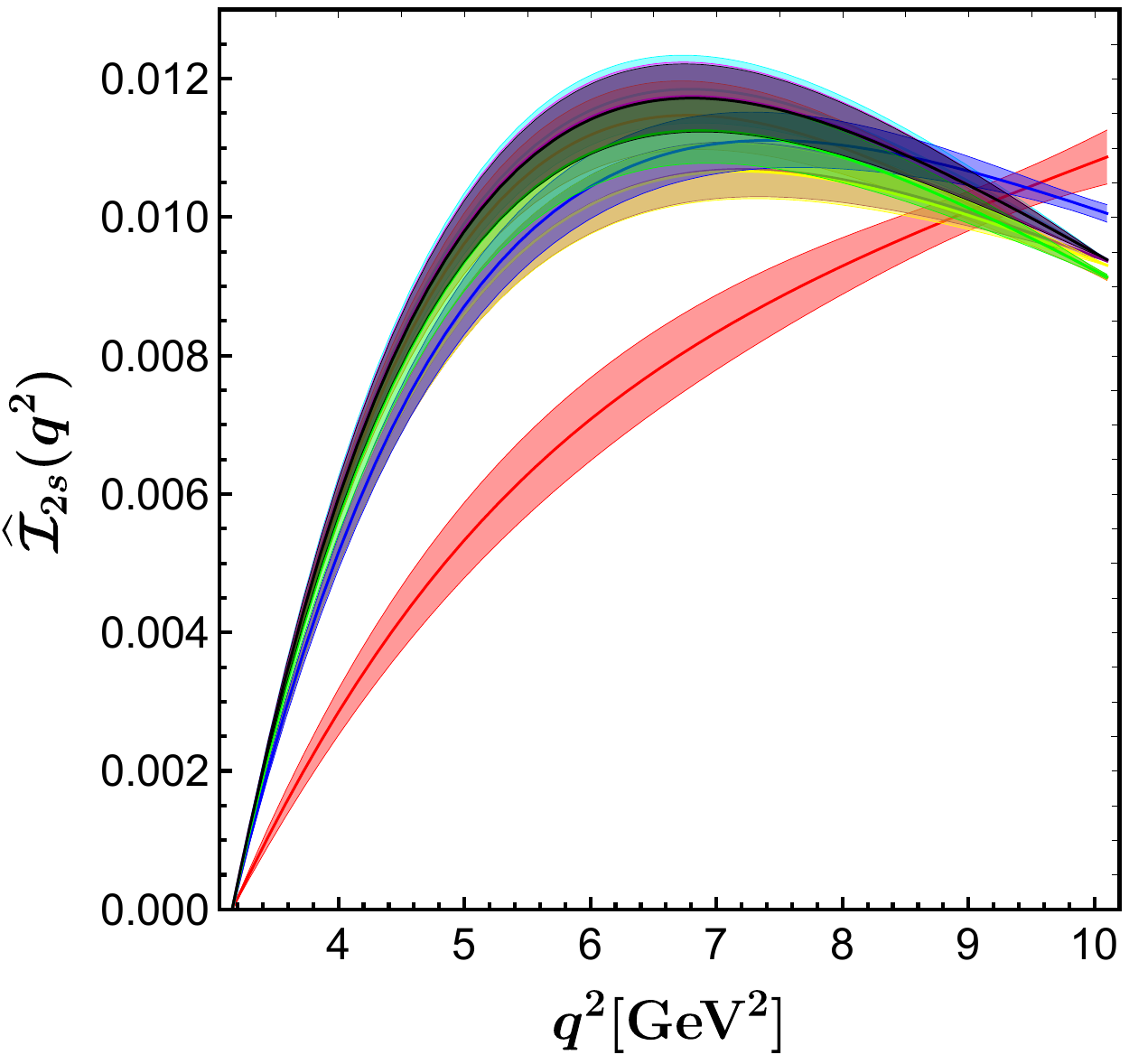}
\includegraphics[width=0.31\textwidth]{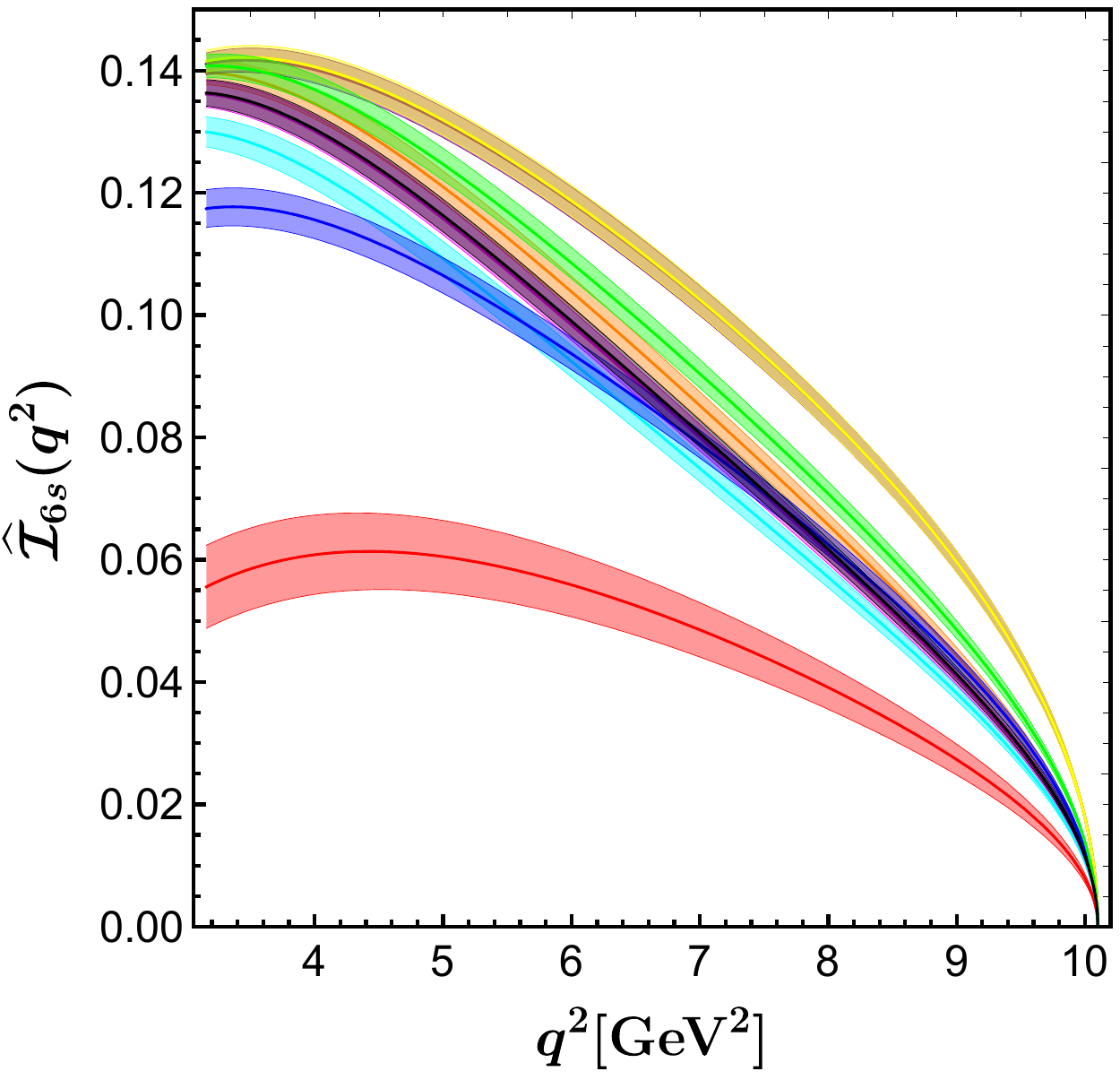}
\\
\includegraphics[width=0.31\textwidth]{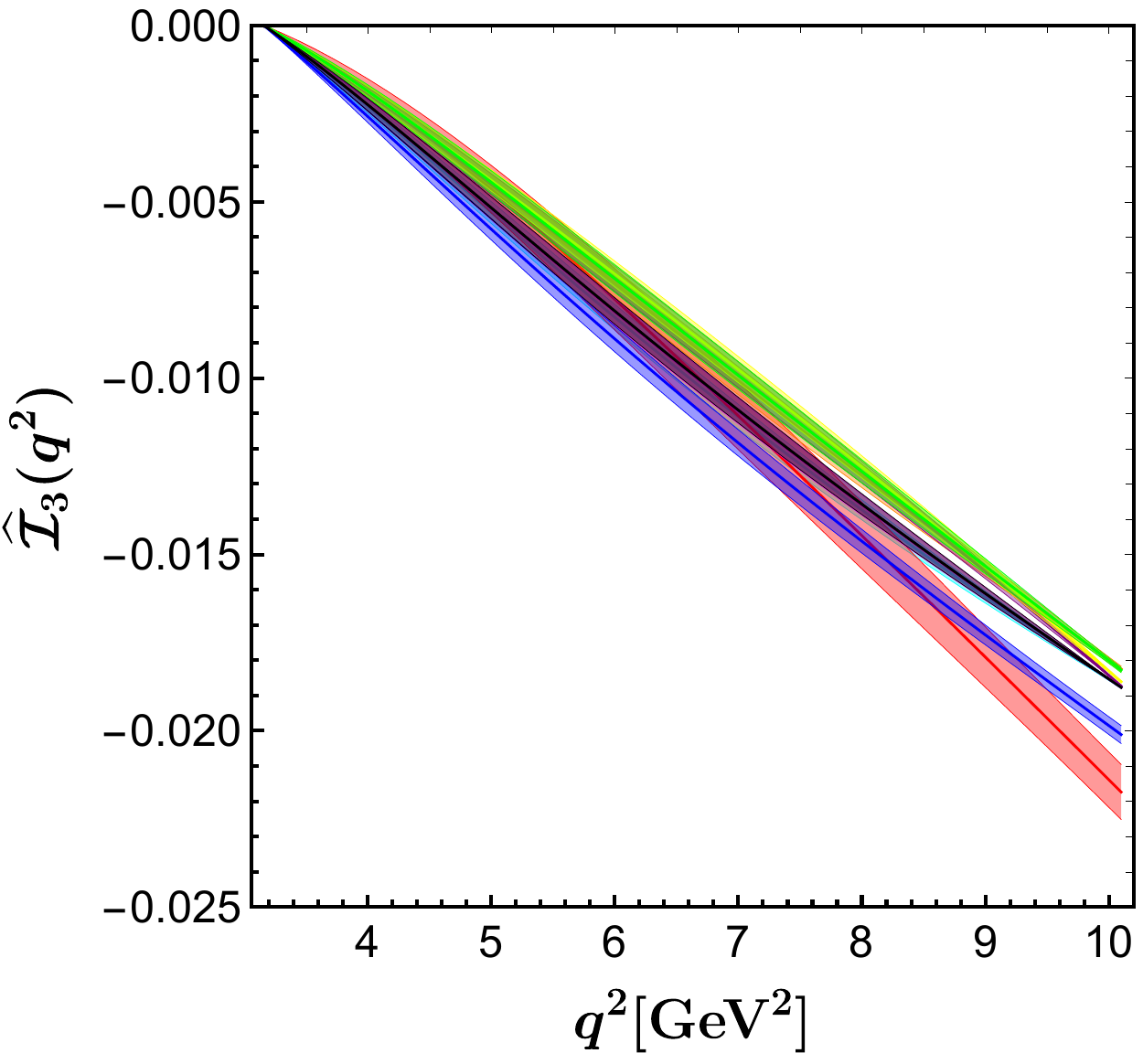}
\includegraphics[width=0.31\textwidth]{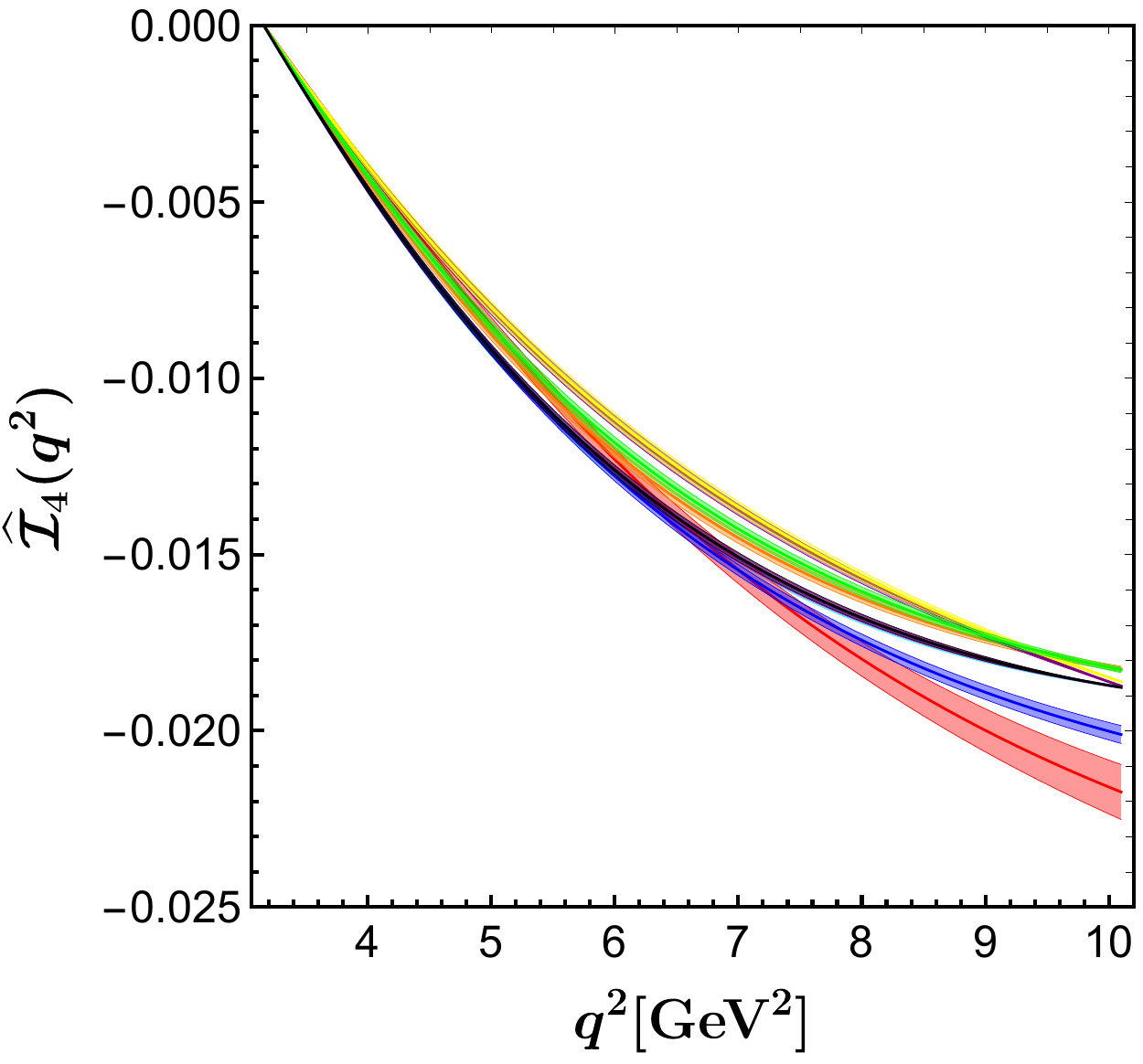}
\includegraphics[width=0.31\textwidth]{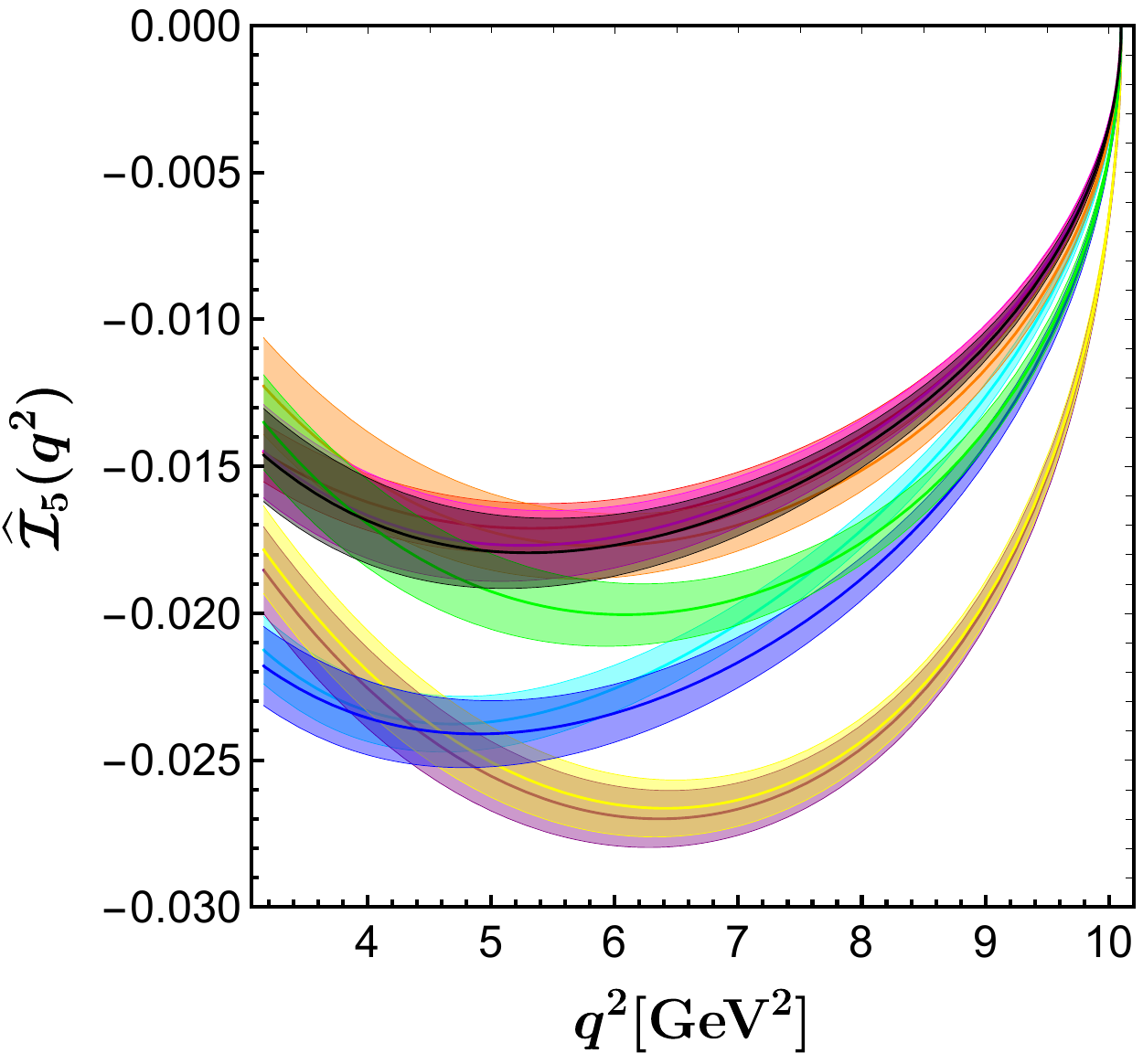}
\\
\includegraphics[width=0.31\textwidth]{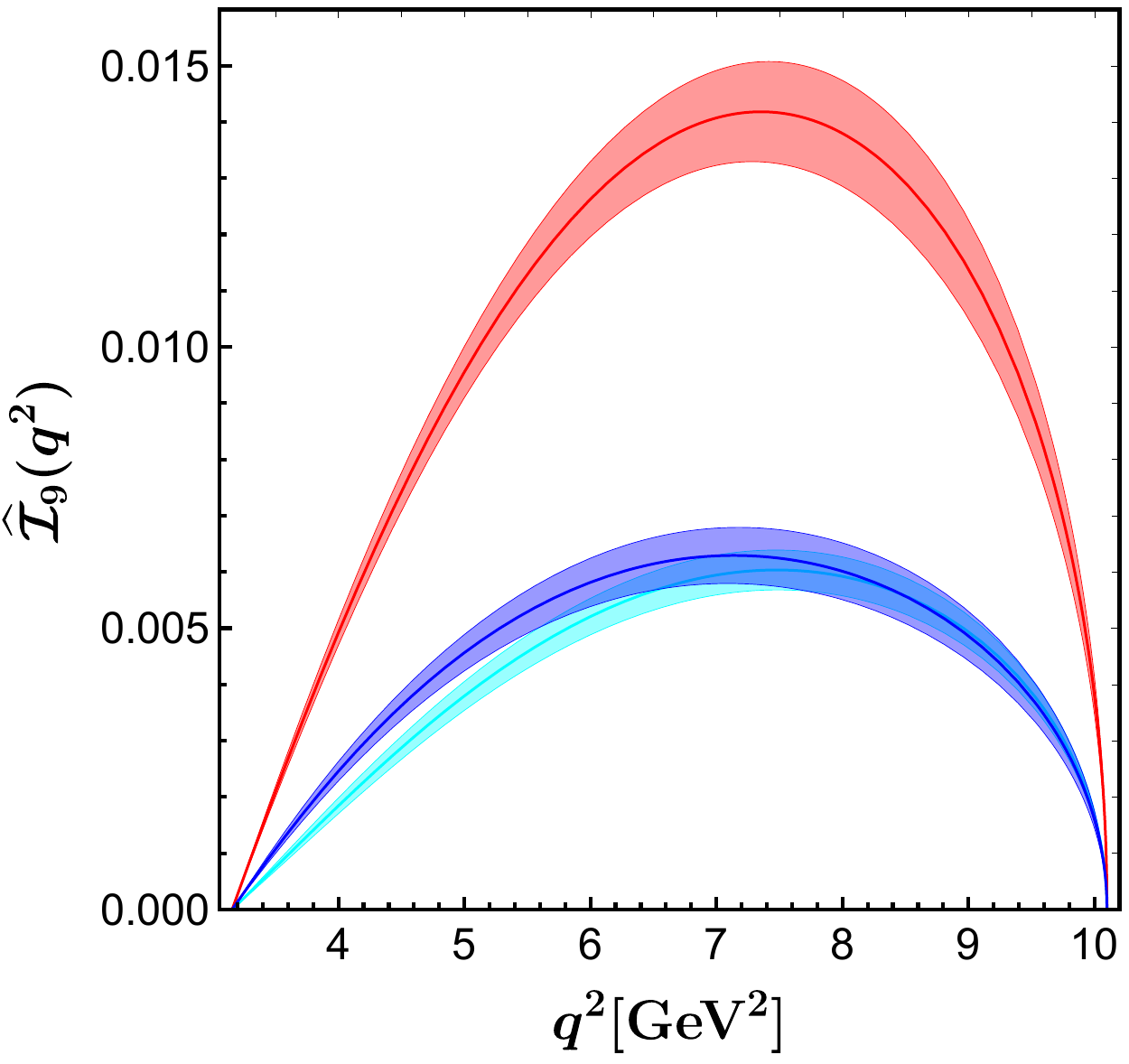}
\includegraphics[width=0.31\textwidth]{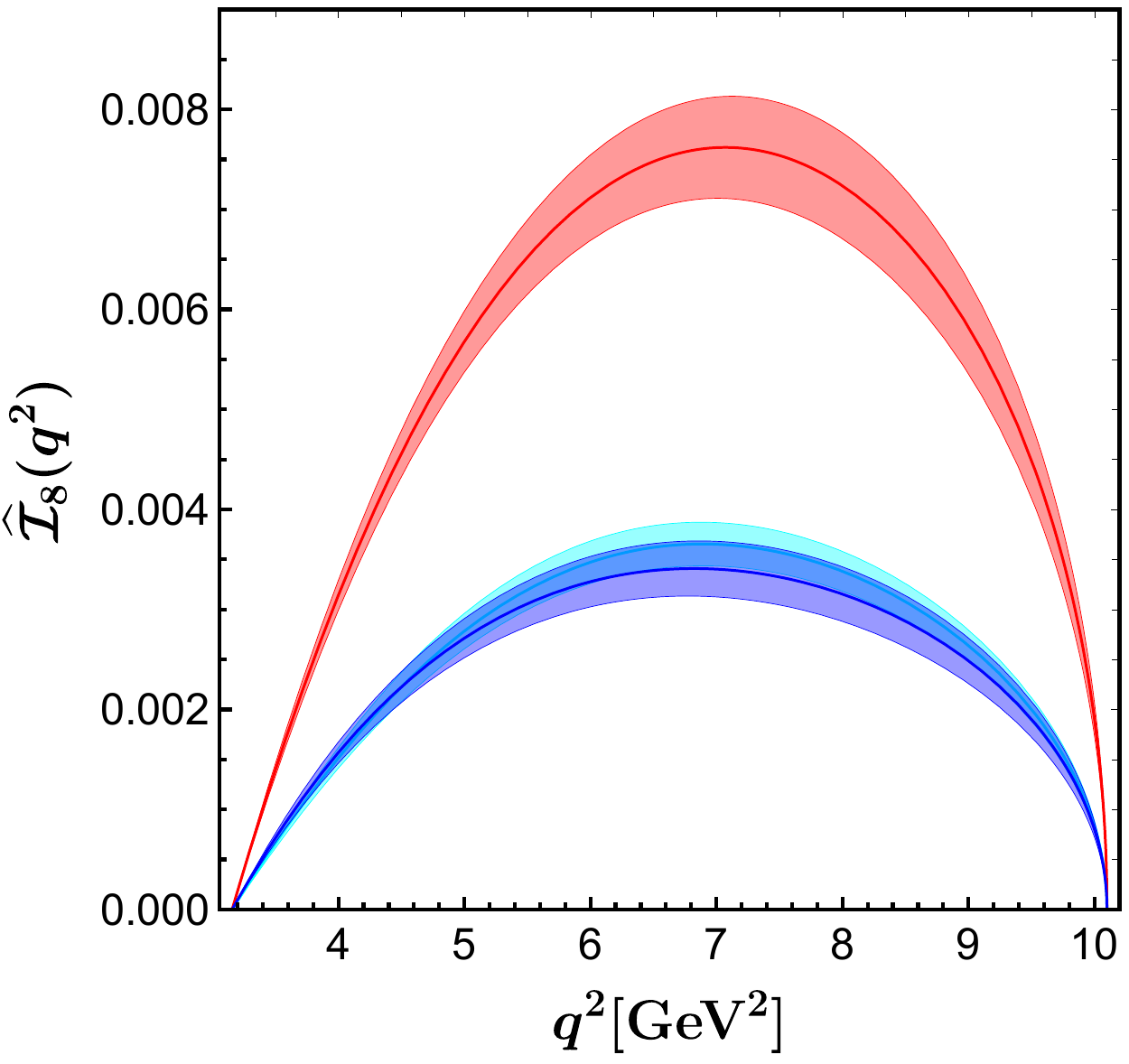}
\includegraphics[width=0.31\textwidth]{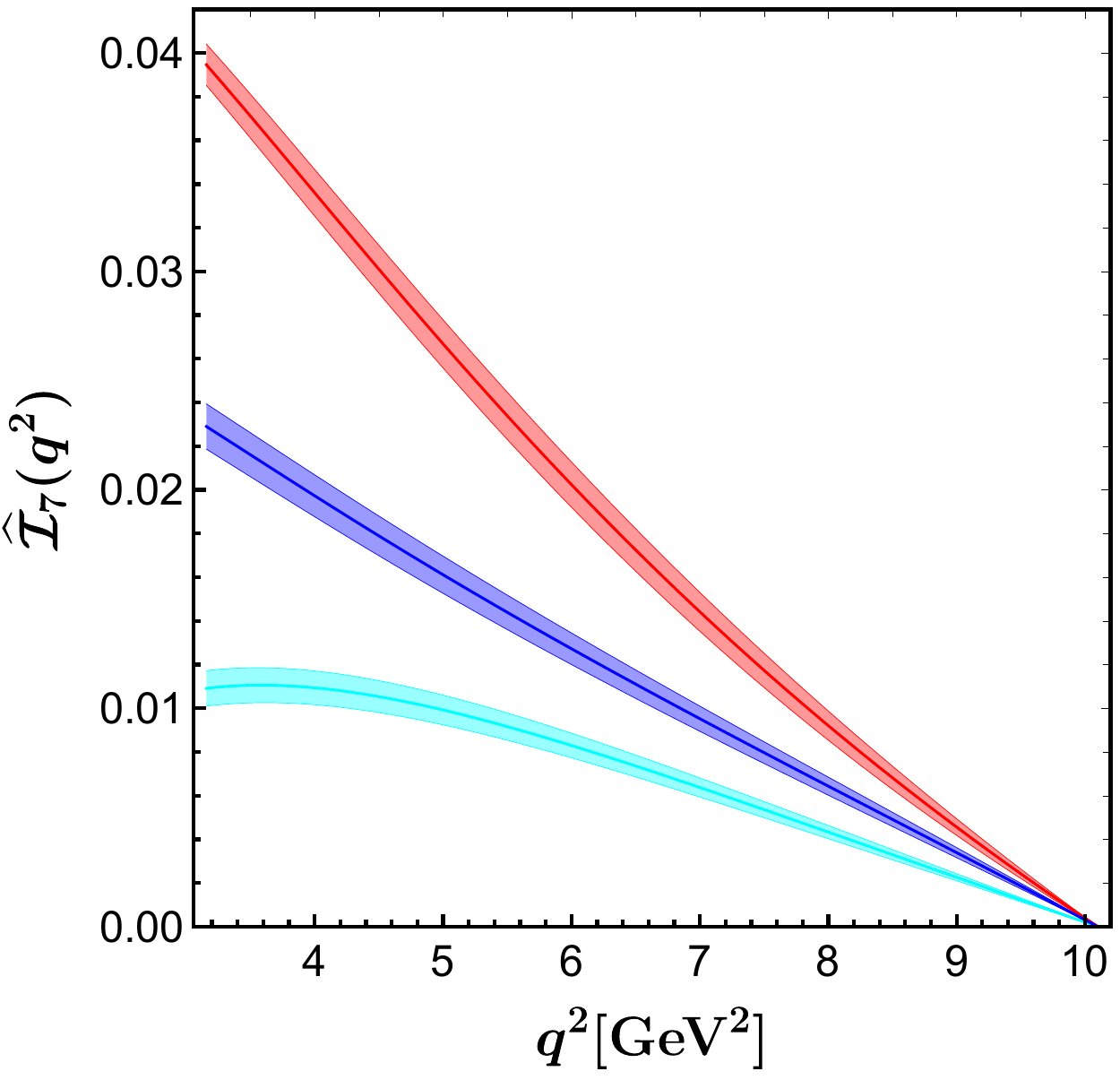}
\caption{\label{fig:angular} \small The angular observables $\widehat{\cal I}_i (q^2)$ as a function of $q^2$, predicted both within the SM and in eight NP benchmark points. The width of each curve is derived from the theoretical uncertainties of $B_c \to J/\psi$ form factors.}
\end{figure}
%%%%%%%%%%%%%%%%%%%%%%%%%%%%%%%%%

In figure~\ref{fig:angular}, we show the predictions for the entire set of angular observables $\widehat{\cal I}_i (q^2)$ within the SM and in eight NP benchmark points. It is easy to see that the BP2 (corresponding to the red band in figure~\ref{fig:angular}) has the greatest effect on all $\widehat{\cal I}_i (q^2)$ except $\widehat{\cal I}_5 (q^2)$. The value of $\widehat{\cal I}_5 (q^2)$ in BP2 is almost the same as that in the SM. The NP corresponding to BP2 even makes the angular observables $\widehat{\cal I}_{6c} (q^2)$ negative, which is not present in the SM and in other NP benchmark points.

In the BP6 (corresponding to the blue band in figure~\ref{fig:angular}), the contributions of NP to all $\widehat{\cal I}_i (q^2)$ except $\widehat{\cal I}_5 (q^2)$ are in the same direction as in BP2, but the impacts are smaller than that in BP2. The BP6 can obviously decrease the value of $\widehat{\cal I}_5 (q^2)$. Observables sensitive to BP6 can be used to study specific UV models, such as the scalar $\mathrm{SU(2)}_L$ doublet $S_2$ (also called $R_2$) leptoquark~\cite{Becirevic:2018afm}, which can produce the relationship $C_{S_L}=4C_T$ at the NP scale.

As we expected, only BP1, BP2, and BP6 which can provide complex phases can produce nonzero angular observables $\widehat{\cal I}_{7,8,9} (q^2)$. The BP1 (corresponding to the cyan band in figure~\ref{fig:angular}) makes $\widehat{\cal I}_{5,6c,6s} (q^2)$ decrease slightly, and hardly contributes to $\widehat{\cal I}_{1c,1s,2c,2s,3,4} (q^2)$. The results of all $\widehat{\cal I}_{i} (q^2)$ predicted by BP4A and BP4B (corresponding to the purple and yellow bands in figure~\ref{fig:angular}, respectively) coincide almost completely with each other. This indicates that $\widehat{\cal I}_{i} (q^2)$ cannot be used to distinguish the two best-fit points of NP hypothesis $\left(C_{S_R},\, C_{S_L} \right) $, which is motivated by models with extra charged Higgs. This is different from the situation in the angular observables of $\Lambda_b^0 \to \Lambda_c^+ (\to \Lambda^0 \pi^+)\tau^- (\to \pi^- \nu_\tau)\bar{\nu}_\tau$ decay, which can distinguish between BP4A and BP4B very well~\cite{Hu:2020axt}. The BP4A and BP4B make $\widehat{\cal I}_{1c,2s,5,6c} (q^2)$ decrease slightly and $\widehat{\cal I}_{1s,2c,3,4,6s} (q^2)$ increase slightly. The NP effects of BP3, BP5, and BP7 have little impact on $\widehat{\cal I}_{i} (q^2)$.

\subsection{Lepton-flavor-universality ratios $R(P_{L,T}^{J/\psi})(q^2)$ and $R(J/\psi)$}

%%%%%%%%%%%%%%%%%%%%%%%%%%%%%%%%%
\begin{figure}[t]
\centering
\includegraphics[width=0.4\textwidth]{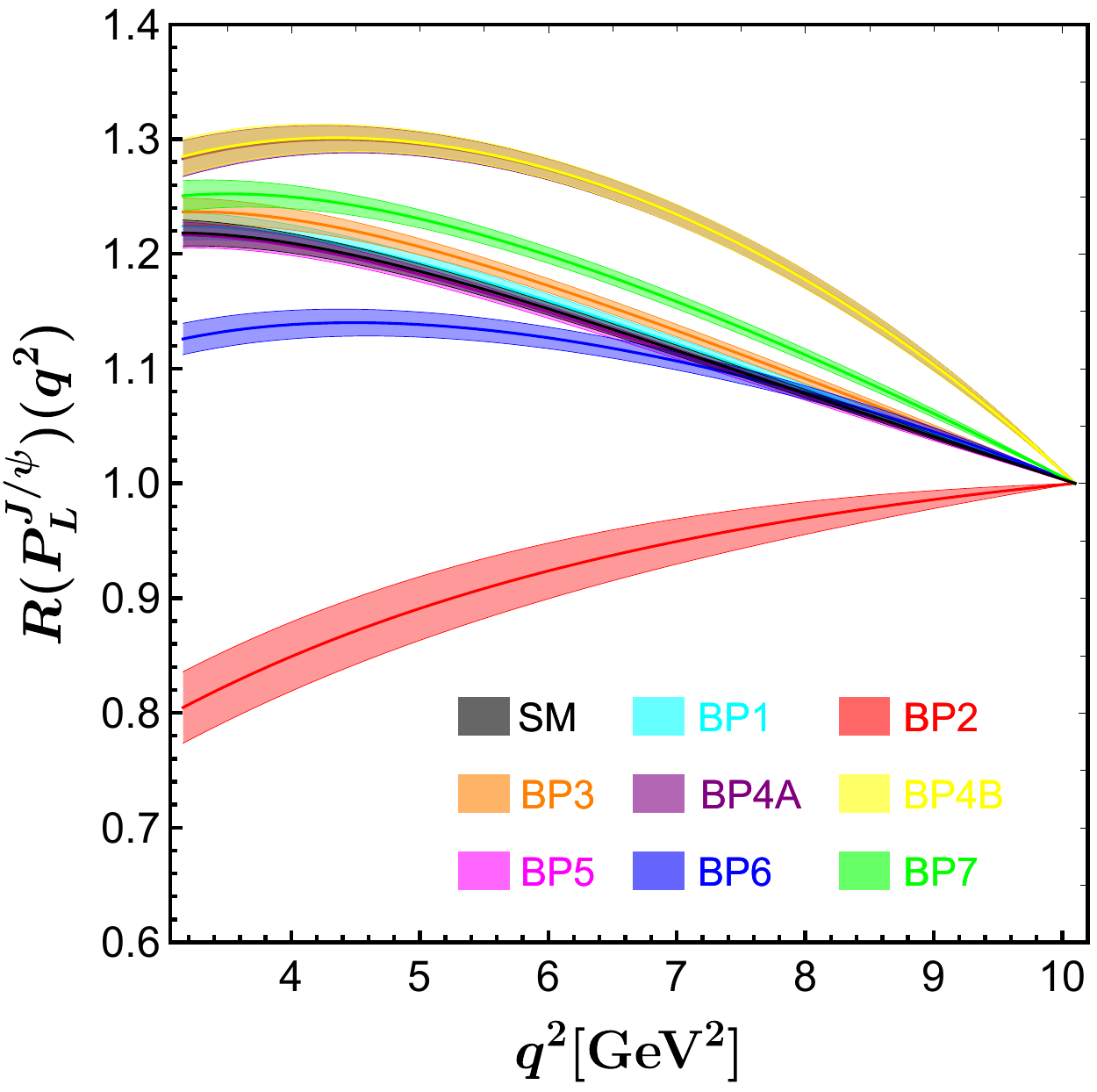}
\quad
\includegraphics[width=0.4\textwidth]{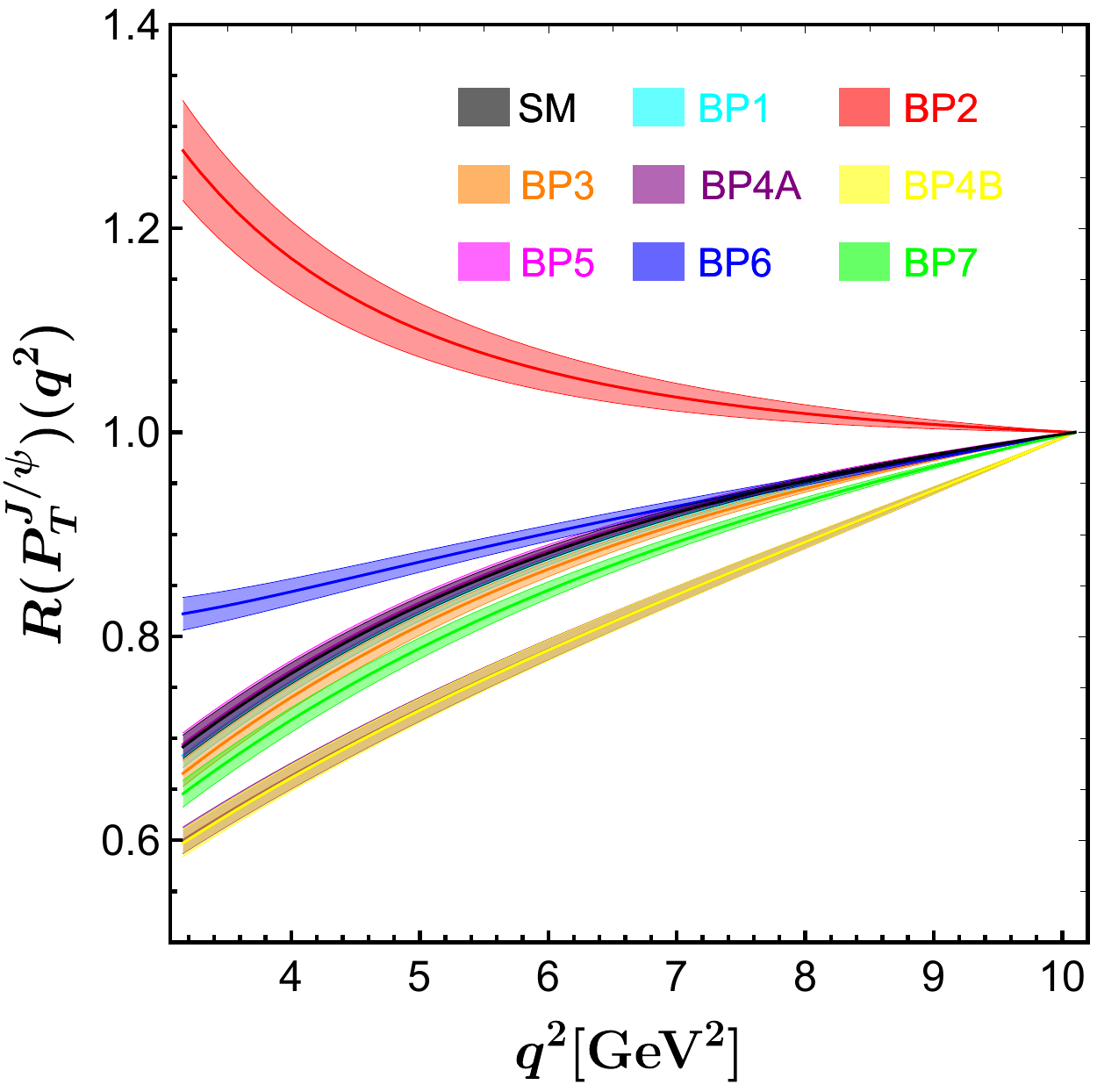}
\caption{\label{fig:RP} \small Lepton-flavor-universality ratios $R(P_{L,T}^{J/\psi})(q^2)$ as a function of $q^2$, predicted both within the SM and in eight NP benchmark points. The width of each curve is derived from the theoretical uncertainties of $B_c \to J/\psi$ form factors.}
\end{figure}
%%%%%%%%%%%%%%%%%%%%%%%%%%%%%%%%%

The $q^2$ distribution of lepton-flavor-universality ratios $R(P_{L,T}^{J/\psi})$ is shown in figure~\ref{fig:RP}, which includes the results within the SM and in eight NP benchmark points. All NP benchmark points except BP5 and BP1 can be distinguished by $R(P_{L,T}^{J/\psi})(q^2)$, especially in the small $q^2$ region. The results of $R(P_{L,T}^{J/\psi})(q^2)$ predicted by BP4A and BP4B coincide almost completely with each other. The longitudinal polarization ratio $R(P_{L}^{J/\psi})(q^2)$ is decreased by benchmarks BP2 and BP6, and increased by benchmarks BP4A, BP4B, BP7, and BP3. Especially, the NP effect of BP2 makes the ratio $R(P_{L}^{J/\psi})(q^2)$ significantly less than 1. The transverse polarization ratio $R(P_{T}^{J/\psi})(q^2)$ is increased by benchmarks BP2 and BP6, and decreased by benchmarks BP4A, BP4B, BP7, and BP3. The NP effect of BP2 makes the ratio $R(P_{T}^{J/\psi})(q^2)$ significantly greater than 1.

All of the NP benchmark points can increase the ratio $R(J/\psi)$. $R(J/\psi)$ does not use the $\tau$ channel for normalization, so the $C_{V_L}$ contribution of BP3, BP5, and BP7 can be seen. The predicted values of $R(J/\psi)$ are shown as follows
\begin{align}
R(J/\psi)_\mathrm{SM} =& 0.2582(38),
&
R(J/\psi)_\mathrm{BP1} =& 0.3272(50),
&
R(J/\psi)_\mathrm{BP2} =& 0.3373(176),
\nonumber\\
R(J/\psi)_\mathrm{BP3} =& 0.3030(45),
&
R(J/\psi)_\mathrm{BP4A} =& 0.2781(36),
&
R(J/\psi)_\mathrm{BP4B} =& 0.2754(36),
\\
R(J/\psi)_\mathrm{BP5} =& 0.3007(44),
&
R(J/\psi)_\mathrm{BP6} =& 0.2938(67),
&
R(J/\psi)_\mathrm{BP7} =& 0.3031(44).\nonumber
\end{align}

\section{Symmetries in the angular observables without tensor operators}
\label{sec:sym}

In the absence of tensor operators, the twelve angular observables $\mathcal{I}_i(q^2,E_{\pi})$ defined in section \ref{sec:Angular distribution} are not independent. These angular observables change to
\begin{align}
\mathcal{I}_{1 c} =&S_1 \left( \left| \mathcal{A}_{\perp}\right|^2 +\left| \mathcal{A}_{\parallel}\right|^2\right) ,\label{eq:I1c_noT}
\\
\mathcal{I}_{1s}=&S_t\left| \mathcal{A}_{t}\right| ^2+(S_1-S_3)\left|\mathcal{A}_{0}\right|^2 
+\frac{1}{2}S_1\left( \left| \mathcal{A}_{\perp}\right|^2+\left| \mathcal{A}_{\parallel}\right|^2\right) , \label{eq:I1s_noT}
\\
\mathcal{I}_{2c}=&S_3 \left( \left| \mathcal{A}_{\perp}\right|^2 +\left| \mathcal{A}_{\parallel}\right|^2\right) , \label{eq:I2c_noT}
\\
\mathcal{I}_{2s}=&-2S_3 \left| \mathcal{A}_{0}\right| ^2+\frac{1}{2}S_3\left(\left| \mathcal{A}_{\perp}\right|^2+ \left| \mathcal{A}_{\parallel}\right|^2\right) , \label{eq:I2s_noT}
\\
\mathcal{I}_{6c}=& 2 S_2 \mathrm{Re}\left[\mathcal{A}_{\parallel}\mathcal{A}_{\perp}^*\right], \label{eq:I6c_noT}
\\
\mathcal{I}_{6s}=&\mathrm{Re}\left[S_2\mathcal{A}_{\parallel}
	\mathcal{A}_{\perp}^*-\sqrt{2}R_t\mathcal{A}_{t} \mathcal{A}_{0}^*\right], \label{eq:I6s_noT}
\\
\mathcal{I}_{3}=&S_3 \left( \left|
	\mathcal{A}_{\perp}\right| ^2- \left|
	\mathcal{A}_{\parallel}\right| ^2\right) , \label{eq:I3_noT}
\\
\mathcal{I}_{9}=& 2 S_3 \mathrm{Im}\left[\mathcal{A}_{\parallel} \mathcal{A}_{\perp}^*\right], \label{eq:I9_noT}
\\
\mathcal{I}_{4}=&\sqrt{2}S_3\mathrm{Re}\left[\mathcal{A}_{\perp} \mathcal{A}_{0}^*\right], \label{eq:I4_noT}
\\
\mathcal{I}_{8}=& \sqrt{2} S_3 \mathrm{Im}\left[\mathcal{A}_{\parallel} \mathcal{A}_{0}^*\right], \label{eq:I8_noT}
\\
\mathcal{I}_{5}=& \frac{1}{2\sqrt{2}}\mathrm{Re}\left[2S_2\mathcal{A}_{\parallel}\mathcal{A}_{0}^* + \sqrt{2}R_t\mathcal{A}_{t}\mathcal{A}_{\perp}^* \right], \label{eq:I5_noT}
\\
\mathcal{I}_{7}=&\frac{1}{2\sqrt{2}}\mathrm{Im}\left[2S_2\mathcal{A}_{\perp}\mathcal{A}_{0}^*-\sqrt{2}R_t\mathcal{A}_{t}\mathcal{A}_{\parallel}^*\right].\label{eq:I7_noT}
\end{align}
We can consider these angular observables as being bilinear in
\begin{equation}
\vec{A}\equiv \left\{\mathrm{Re}[\mathcal{A}_{t}],\mathrm{Im}[\mathcal{A}_{t}], \mathrm{Re}[\mathcal{A}_{0}],\mathrm{Im}[\mathcal{A}_{0}],\mathrm{Re}[\mathcal{A}_{\perp}],\mathrm{Im}[\mathcal{A}_{\perp}],\mathrm{Re}[\mathcal{A}_{\parallel}],\mathrm{Im}[\mathcal{A}_{\parallel}]\right\}.
\end{equation}
Generally, the experimental and theoretical degrees of freedom can be matched by the following formula~\cite{Egede:2010zc,Matias:2012xw,Hofer:2015kka,Alguero:2020ukk}
\begin{equation}
n_c - n_d = 2n_A - n_s,
\end{equation}
where $n_c$ is the number of angular observables $\mathcal{I}_i$; $n_d$ is  the number of dependencies between the different observables $\mathcal{I}_i$, which can be obtained by the difference between the number of observables $\mathcal{I}_i$ and the dimension of the space given by the gradient vectors $\vec{\nabla}\mathcal{I}_i$ (with the derivatives taken with respect to the various elements of $\vec{A}$); $n_A$ is the number of transversity amplitudes (each $\mathcal{A}_{j}$ is complex and therefore has two degrees of freedom); $n_s$ is the number of continuous symmetries.

Without tensor operators, there are still twelve angular observables $\mathcal{I}_i$ but only four amplitudes $\mathcal{A}_{t,0,\perp,\parallel}$. So $n_c = 12$ and $n_A = 4$. In this case, the only continuous symmetry that can be found is
\begin{equation}
\label{eq:conti_sym}
\mathcal{A}_{t}\to e^{i\alpha} \mathcal{A}_{t},
\quad
\mathcal{A}_{0}\to e^{i\alpha} \mathcal{A}_{0},
\quad
\mathcal{A}_{\perp}\to e^{i\alpha} \mathcal{A}_{\perp},
\quad
\mathcal{A}_{\parallel}\to e^{i\alpha} \mathcal{A}_{\parallel}.
\end{equation}
Only 7 of the 12 angular observables $\mathcal{I}_i$ are independent and 5 dependencies are found. We present the dependence relations directly here and provide the detailed derivation in appendix~\ref{sec:dep}:
\begin{align}
S_1 \mathcal{I}_{2c}=& S_3 \mathcal{I}_{1c},\label{eq:relation1}
\\
S_2^2 \mathcal{I}_{2c}^2=& S_2^2 \left(\mathcal{I}_{3}^2+\mathcal{I}_{9}^2\right)+ S_3^2 \mathcal{I}_{6c}^2,\label{eq:relation2}
\\
4 S_2^2 \beta_2^2=& S_3^2 \mathcal{I}_{6c}^2 \left[\left(\mathcal{I}_{2c}-2\mathcal{I}_{2s}\right) \left(\mathcal{I}_{2c}+\mathcal{I}_{3}\right)-4\mathcal{I}_{4}^2\right],\label{eq:relation3}
\\
R_t^2\left(\mathcal{I}_{2c} -2 \mathcal{I}_{2s} \right)& \left[4 S_3 \mathcal{I}_{1s} + \left(S_3-3 S_1\right) \mathcal{I}_{2c} +2 \left(S_1-S_3\right) \mathcal{I}_{2s}\right]\nonumber
\\
=&2 S_t\left\{S_3^2 \left(\mathcal{I}_{6c}-2 \mathcal{I}_{6s}\right)^2 +\frac{\left[S_3^2 \mathcal{I}_{6c} \beta_3 + S_2^2  \left(\mathcal{I}_{2c}-2 \mathcal{I}_{2s}\right) \mathcal{I}_{9} \beta_2\right]^2}{S_2^2 \left(\mathcal{I}_{2c}+\mathcal{I}_{3}\right)^2 \beta_2^2}\right\},\label{eq:relation4}
\\
2S_3^2 \left(\mathcal{I}_{2c}+\mathcal{I}_{3}\right)^2 \mathcal{I}_{7}
=&
S_2^2 \frac{\left[(\mathcal{I}_{2c}+\mathcal{I}_{3})^2+\mathcal{I}_{9}^2\right] \beta_2}{\mathcal{I}_{6c}}- \frac{S_3^4 \mathcal{I}_{4} \mathcal{I}_{6c}^2 \beta_3}{S_2^2 \left(\mathcal{I}_{2c} -2 \mathcal{I}_{2s} \right)\beta_2}\nonumber
\\
&+S_3^2\frac{\mathcal{I}_{9} \left[\beta_3 -\left(\mathcal{I}_{2c}-2\mathcal{I}_{2s}\right) \mathcal{I}_{4} \mathcal{I}_{6c}\right] +\mathcal{I}_{8} \left(\mathcal{I}_{2c}+\mathcal{I}_{3}\right)^2 \left(\mathcal{I}_{6c}-2 \mathcal{I}_{6s}\right)}{\mathcal{I}_{2c}-2 \mathcal{I}_{2s}},\label{eq:relation5}
\end{align}
with
\begin{align}
\beta_2 \equiv & \left(\mathcal{I}_{2c}+\mathcal{I}_{3}\right) \mathcal{I}_{8} - \mathcal{I}_{4} \mathcal{I}_{9},\label{eq:beta2}
\\
\beta_3\equiv & 2\mathcal{I}_{2c} \mathcal{I}_{5} \left(\mathcal{I}_{2c}-2\mathcal{I}_{2s} + \mathcal{I}_{3}\right) +2\mathcal{I}_{4}\mathcal{I}_{6s} \left(\mathcal{I}_{2c}+\mathcal{I}_{3}\right) \nonumber
\\
&+\mathcal{I}_{4} \mathcal{I}_{6c} \left(2 \mathcal{I}_{2s}-\mathcal{I}_{3}-2\mathcal{I}_{2c}\right) -4 \mathcal{I}_{2s} \mathcal{I}_{3} \mathcal{I}_{5}.\label{eq:beta3}
\end{align}
Eqs.~\eqref{eq:relation1}--\eqref{eq:relation5} can be used as a model-independent method to determine the existence of tensor operators. The ``model-independent method" here not only means that it does not depend on the NP models, but also means that it does not depend on the calculation of $B_c \to J/\psi$ transition form factors. 

Furthermore, we can obtain the dependence relations among the normalized angular observables $\widehat{\mathcal{I}}_i(q^2)$ by replacing the $\mathcal{I}_i(q^2,E_\pi)$ and the dimensionless factors $S_{t,1,2,3}$ and $R_t$ in eqs.~\eqref{eq:relation1}--\eqref{eq:beta3} with $\widehat{\mathcal{I}}_i(q^2)$, $\bar{S}_{t,1,2,3}$ and $\bar{R}_t$, respectively. The factors $\bar{S}_{t,1,2,3}$ and $\bar{R}_t$ are defined, respectively, as
\begin{align}
\bar{S}_{t,1,2,3} \equiv \int{S_{t,1,2,3}} dE_\pi,
\quad
\bar{R}_t \equiv \int{R_t} dE_\pi.
\end{align}

\section{Conclusions}
\label{sec:conclusions}

Inspired by the $R(D^{(*)})$ anomalies, the angular distribution of $B_c^- \to J/\psi \tau^-\bar{\nu}_\tau$ or $B_c^- \to J/\psi (\to \mu^+ \mu^-)\tau^-\bar{\nu}_\tau$ decay has been used to explore possible NP patterns in $b \to c \tau^- \bar{\nu}_\tau$ transition in many previous works. However, angular observables depending on the solid angle of final-state $\tau^-$ are unmeasurable theoretically, since the decay products of $\tau^-$ inevitably contain an undetected $\nu_\tau$ and the solid angle of $\tau^-$ cannot be determined precisely. Therefore, in this work, we study the {\it measurable} angular distribution of the five-body decay $B_c^- \to J/\psi (\to \mu^+ \mu^-)\tau^- (\to \pi^- \nu_\tau)\bar{\nu}_\tau$, which includes three visible final-state particles $\mu^+$, $\mu^-$, and $\pi^-$, with their three-momenta all being measured. 

The five-fold differential decay rate containing all NP effective operators can be expressed in terms of twelve angular observables $\mathcal{I}_{i} (q^2,E_{\pi})$, which can be completely expressed by seven independent transversity amplitudes and some dimensionless factors. As long as one of the angular observables $\mathcal{I}_{7}$, $\mathcal{I}_{8}$ and $\mathcal{I}_{9}$ is nonzero, this will be an unquestionable sign of NP, and indicates that the NP can cause extra weak phases. Integrating the five-fold differential decay rate over the $E_\pi$ and normalized by $d\Gamma/dq^2$, we can construct twelve normalized angular observables $\widehat{\mathcal{I}}_{i} (q^2)$. By integrating over all lepton-side parameters, we find that there are only two angular observables $P^{J/\psi}_{L,T}(q^2)$ whose determination can be obtained without reconstruction of the dilepton solid angle. The $P^{J/\psi}_{L,T}(q^2)$ are not affected by the lepton dynamics, so they can be used to construct the ratios $R(P^{J/\psi}_{L,T})$ to probe the universality of lepton flavor. Based on our five-fold differential decay rate, we show how to extract the complete set of $\tau$ asymmetries in $B_c^- \to J/\psi \tau^-\bar{\nu}_\tau$ decay from the visible final-state kinematics.

Using the $B_c \to J/\psi$ vector and axial-vector form factors calculated by the latest lattice QCD and the tensor form factors calculated by the QCD sum rule, we predict the $q^2$ distribution of the twelve normalized angular observables $\widehat{\mathcal{I}}_{i}$ and the two lepton-flavor-universality ratios $R(P_{L,T}^{J/\psi})$ both within the SM and in eight NP benchmark points, which are a variety of best-fit points in seven different NP hypotheses. We find that the benchmark BP2 (corresponding to the hypothesis of tensor operator) has the greatest effect on all $\widehat{\mathcal{I}}_{i}$ and $R(P_{L,T}^{J/\psi})$, except $\widehat{\mathcal{I}}_{5}$. Especially for the observables $\widehat{\mathcal{I}}_{6c}$ and $R(P_{L,T}^{J/\psi})$, BP2 makes their predictions very different from those in the SM and other benchmark points. The results of all $\widehat{\mathcal{I}}_{i}$ and $R(P_{L,T}^{J/\psi})$ predicted by the two best-fit points (i.e. BP4A and BP4B) of NP hypothesis $(C_{S_R},\, C_{S_L})$, which is motivated by models with extra charged Higgs, coincide almost completely with each other. This is different from the situation in the angular observables of the baryonic counterparts, which can distinguish between BP4A and BP4B very well. In addition to the benchmarks BP2, BP4A, and BP4B, the BP1, BP6, and BP7 can also have some influence on the observables. Compared with the $\widehat{\mathcal{I}}_{i}$, the ratios $R(P_{L,T}^{J/\psi})$ are more sensitive to the NP with pseudo-scalar operator. All NP benchmark points can improve the value of $R(J/\psi)$, which makes it closer to the experimental measurement.

We discuss the symmetries in the angular observables without tensor operators, and present five dependence relations. Once all twelve angular observables are measured, these five relations will be a very useful way to determine the existence of tensor operators. If these relations are not fulfilled, it means that there must be tensor operators. This method is completely independent of any assumptions on the details of the NP model and $B_c \to J/\psi$ transition form factors. 

The cascade decay $B_c^- \to J/\psi (\to \mu^+ \mu^-)\tau^- (\to \pi^- \nu_\tau)\bar{\nu}_\tau$ provides a good prospect for measurement by the LHCb experiment, because it has excellent final-state signatures with a strongly peaking $\mu^+\mu^-$ spectrum and a well $\pi^-$ identification. Additionally, the lifetime of $B_c$ meson is almost three times shorter than that of $B_{u,d,s}$ mesons, which can be used to improve the separation of $B_c$ decay from $B_{u,d,s}$ decays, thus providing an extra handle to distinguish the large background that derives from the $B_{u,d,s}$ mesons~\cite{Aaij:2017tyk,Bernlochner:2021vlv}. One possible background is the very rare $B^- \to \pi^- \mu^+ \mu^-$ decay~\cite{Aaij:2015nea}, which has about one-tenth the number of events as the signal decay. This background can also be distinguished from the signal by the kinematic properties of the visible final-state particles. Future precise measurements of the angular observables in $B_c^- \to J/\psi (\to \mu^+ \mu^-)\tau^- (\to \pi^- \nu_\tau)\bar{\nu}_\tau$ decay, especially precise measurements of the normalized ones, would be very helpful to provide a more definite answer concerning the anomalies observed in $b \to c \tau^- \bar{\nu}_\tau$ transition, restricting further or even deciphering the NP models.

\acknowledgments

This work is supported by the National Natural Science Foundation of China under Grant Nos.~11947083, 12075097, 11675061 and 11775092, as well as by the CCNU-QLPL Innovation Fund (QLPL2020P01). X.L. is also supported by the Fundamental Research Funds for the Central Universities under Grant No.~CCNU20TS007.

\appendix

\section{The calculation of the angular distribution}
\label{sec:ang_dis}

In the appendix of ref.~\cite{Hu:2020axt}, we have given the detailed calculation procedures for the similar five-body cascade decay of unpolarized $\Lambda_b$ baryon. The calculation of $V^* \to \tau^- (\to \pi^- \nu_\tau)\bar{\nu}_\tau$ part is exactly the same as that in this work. Therefore, in this section, we mainly present some important definitions and conventions, and calculate the $J/\psi \to \mu^+ \mu^-$ decay. At the end of this section, the dimensionless factors are listed for the sake of completeness of this paper.

\subsection{Definitions and conventions}

The differential decay rate of cascade decay $B_c^- \to J/\psi (\to \mu^+ \mu^-)\tau^- (\to \pi^- \nu_\tau)\bar{\nu}_\tau$ can be written as 
\begin{align}
d\Gamma =& 
\frac{4 G_F^2 \left|V_{cb} \right|^2 {\cal B}\left(\tau \to \pi^- \nu_\tau \right)  dq^2 }{ m_{B_c} \left(m_\tau^2 - m_\pi^2 \right)^2  m_{J/\psi} \Gamma_{J/\psi}} \sum_{\lambda_{\mu^-},\lambda_{\mu^+}} \left|{\sf M}_{\lambda_{\mu^-},\lambda_{\mu^+}} \right|^2 
d\Pi_2\left(p_{B_c}; q, p_{J/\psi} \right) 
\nonumber \\
& \times 
d\Pi_2\left(q; p_\tau, p_{\bar{\nu}} \right) 
d\Pi_2\left(p_\tau; p_{\pi^-}, p_\nu \right) 
d\Pi_2\left(p_{J/\psi}; p_{\mu^-}, p_{\mu^+} \right),
\end{align}
where 
\begin{align}
{\sf M}_{\lambda_{\mu^-},\lambda_{\mu^+}} 
&\equiv 
{\sf H}_{\lambda_{\mu^-},\lambda_{\mu^+}} {\sf L} 
+ \sum_{\lambda} \eta_\lambda {\sf H}_{\lambda_{\mu^-},\lambda_{\mu^+}}^\lambda {\sf L}_\lambda
+ \sum_{\lambda, \lambda'} \eta_\lambda \eta_{\lambda^\prime} {\sf H}_{\lambda_{\mu^-},\lambda_{\mu^+}}^{\lambda, \lambda'} {\sf L}_{\lambda, \lambda'},
\\
{\sf H}_{\lambda_{\mu^-},\lambda_{\mu^+}} 
&= 
\sum_{\lambda_{J/\psi}} H_{\lambda_{J/\psi}} 
{\cal M}^{\lambda_{J/\psi}}_{\lambda_{\mu^-},\lambda_{\mu^+}},
\\
{\sf H}_{\lambda_{\mu^-},\lambda_{\mu^+}}^\lambda 
&= 
\sum_{\lambda_{J/\psi}} H_{\lambda_{J/\psi}}^\lambda 
{\cal M}^{\lambda_{J/\psi}}_{\lambda_{\mu^-},\lambda_{\mu^+}},
\\
{\sf H}_{\lambda_{\mu^-},\lambda_{\mu^+}}^{\lambda,\lambda'} 
&= 
\sum_{\lambda_{J/\psi}} H_{\lambda_{J/\psi}}^{\lambda,\lambda'} 
{\cal M}^{\lambda_{J/\psi}}_{\lambda_{\mu^-},\lambda_{\mu^+}}.
\end{align}
The $\lambda_{\mu^-}$ and $\lambda_{\mu^+}$ respectively represent the helicity of final-state particles $\mu^-$ and $\mu^+$, as well as the $\lambda^{(\prime)}$ and $\lambda_{J/\psi}$ respectively represent the helicity of intermediate state $V^*$ and $J/\psi$. The $d\Pi_2$ stands for the two-body phase space. ${\cal M}^{\lambda_{J/\psi}}_{\lambda_{\mu^-},\lambda_{\mu^+}}$ denotes the helicity amplitude related to $J/\psi \to \mu^+ \mu^-$ decay. $H_{\lambda_{J/\psi}}^{\left(\lambda, \lambda'\right) }$ are the hadronic helicity amplitudes describing the $B_c \to J/\psi$ transition with different Lorentz structures. 
\begin{align}
H_{\lambda_{J/\psi}} & =  \left\langle J/\psi(\lambda_{J/\psi}) \left|g_S ({\bar c} b) + g_P ({\bar c} \gamma_5 b) \right| B_c \right\rangle, 
\\
H^{\lambda}_{\lambda_{J/\psi}} & =  \epsilon^{\mu*}(\lambda)\left\langle J/\psi(\lambda_{J/\psi}) \left|g_V ({\bar c}\gamma_\mu b) + g_A ({\bar c}\gamma_\mu \gamma_5 b) \right|B_c \right\rangle, 
\\
H^{\lambda,\lambda'}_{\lambda_{J/\psi}} & =  g_T \epsilon^{\mu*}(\lambda) \epsilon^{\nu*}(\lambda') \left\langle J/\psi(\lambda_{J/\psi}) \left| {\bar c} i \sigma_{\mu\nu}(1-\gamma_5) b \right|B_c \right\rangle, 
\end{align}
where $\epsilon^\mu(\lambda)$ denotes the polarization vector of the virtual vector boson $V^*$ with helicity $\lambda$. The modified leptonic helicity amplitudes are ${\sf L}_{\left(\lambda, \lambda' \right) }$, which can be obtained directly from the appendix of ref.~\cite{Hu:2020axt}.

In the $B_c$ rest frame, the polarization vector of $J/\psi$ meson can be written as~\cite{Auvil:1966eao,Haber:1994pe} 
\begin{align}
\varepsilon^\mu\left(\pm 1 \right) &= \left(0,\, \mp 1, \, -i, \, 0 \right)/\sqrt{2},
\\
\varepsilon^\mu\left( 0 \right) &= \left(\left|{\bm p}_{J/\psi} \right|, \, 0, \, 0,\, E_{J/\psi}  \right) /m_{J/\psi},
\end{align}
with $\left|{\bm p}_{J/\psi} \right| = \sqrt{Q_+ Q_-}/\left(2 m_{B_c} \right)$ and $E_{J/\psi} = \left(m_{B_c}^2 + m_{J/\psi}^2 - q^2 \right) /\left(2 m_{B_c} \right)$. The polarization vector of virtual $V^*$ can be written as~\cite{Auvil:1966eao,Haber:1994pe}
\begin{align}
\epsilon^\mu\left(\pm 1 \right) & = \left(0,\, \pm1,\, -i,\, 0 \right)/\sqrt{2}, \label{eq:polV1}
\\
\epsilon^\mu\left(0 \right) & = \left(\left|{\bm q} \right|,\, 0,\, 0,\, -q_0  \right) /\sqrt{q^2}, \label{eq:polV0}
\\
\epsilon^\mu\left(t \right) & = q^\mu/\sqrt{q^2}, \label{eq:polVt}
\end{align}
with $\left|{\bm q}\right| = \sqrt{Q_+ Q_-}/\left(2 m_{B_c} \right)$ and $q_0 = \left(m_{B_c}^2 - m_{J/\psi}^2 + q^2 \right) /\left(2 m_{B_c} \right)$. Eqs.~\eqref{eq:polV1}--\eqref{eq:polVt} satisfy the completeness relation
\begin{equation}
\label{eq:completeness_relation}
g^{\mu\nu}=\sum_{\lambda \in \{t,\pm1,0\}} \epsilon^{\mu}(\lambda) \epsilon^{\nu*}(\lambda) \eta_{\lambda},
\end{equation}
where $\eta_{t} = 1$ and $\eta_{\pm1,0} = -1$.

For scalar and pseudo-scalar operators, there is only one nonzero hadronic helicity amplitude
\begin{equation}
H_0 = {\cal A}_t^{SP}.
\end{equation}

For vector and axial-vector operators, there are four nonzero hadronic helicity amplitudes listed as follows
\begin{align}
H_0^t &= {\cal A}_t^{VA}, &
H_0^0 &= {\cal A}_0, 
\nonumber \\
H_1^1 &= \left({\cal A}_\perp + {\cal A}_\parallel \right)/\sqrt{2}, &
H_{-1}^{-1} &= \left({\cal A}_\perp - {\cal A}_\parallel \right)/\sqrt{2}. 
\end{align}

For the tensor operators, there are twelve nonzero hadronic helicity amplitudes listed as follows
\begin{align}
H_0^{t,0} &= H_0^{-1,1} = -H_0^{0,t} = -H_0^{1,-1} = {\cal A}_0^T, 
\nonumber \\
H_1^{0,1} &= H_1^{t,1} = -H_1^{1,0} = -H_1^{1,t} = \left({\cal A}_\parallel^T + {\cal A}_\perp^T \right)/\sqrt{2},
\nonumber \\
H_{-1}^{0,-1} &= H_{-1}^{-1,t} = -H_{-1}^{-1,0} = -H_{-1}^{t,-1} = \left({\cal A}_\parallel^T - {\cal A}_\perp^T \right)/\sqrt{2}.
\end{align}

\subsection{Calculating $J/\psi \to \mu^+ \mu^-$ decay}

The $J/\psi \to \mu^+ \mu^-$ decay should be calculated in the $J/\psi$ rest frame. In this reference frame, the transverse polarization vector of $J/\psi$ meson does not change, i.e., $\tilde{\varepsilon}^\mu\left(\pm1 \right) = \varepsilon^\mu\left(\pm1 \right)$, but its longitudinal polarization vector changes to $\tilde{\varepsilon}^\mu\left(0 \right) = \left(0,\, 0,\, 0,\, 1 \right) $. For massless $\mu^-$ and $\mu^+$ leptons, their spinors can be written as~\cite{Auvil:1966eao,Haber:1994pe}
\begin{align}
	\label{eq:muspinors}
	u\left(\frac{1}{2}\right)&=\sqrt{\frac{m_{J/\psi}}{2}}\left(\cos \frac{\theta_{J/\psi}}{2}, \ \sin \frac{\theta_{J/\psi}}{2},\ \cos \frac{\theta_{J/\psi}}{2}, \ \sin \frac{\theta_{J/\psi}}{2}\right)^T,
	\\
	u\left(-\frac{1}{2}\right)&=\sqrt{\frac{m_{J/\psi}}{2}}\left(-\sin \frac{\theta_{J/\psi}}{2},\ \cos \frac{\theta_{J/\psi}}{2},\ \sin \frac{\theta_{J/\psi}}{2},\ -\cos \frac{\theta_{J/\psi}}{2}\right)^T,
	\\
	v\left(\frac{1}{2}\right)&=\sqrt{\frac{m_{J/\psi}}{2}} \left(\cos \frac{\theta_{J/\psi}}{2}, \ \sin \frac{\theta_{J/\psi}}{2},\ -\cos \frac{\theta_{J/\psi}}{2}, \ -\sin \frac{\theta_{J/\psi}}{2}\right)^T,
	\\
	v\left(-\frac{1}{2}\right)&=\sqrt{\frac{m_{J/\psi}}{2}} \left(-\sin \frac{\theta_{J/\psi}}{2},\ \cos \frac{\theta_{J/\psi}}{2},\ -\sin \frac{\theta_{J/\psi}}{2},\ \cos \frac{\theta_{J/\psi}}{2}\right)^T,
\end{align}
where the Jacob-Wick second particle convention has been used~\cite{Jacob:1959at}.

As the $J/\psi \to \mu^+\mu^-$ decay is dominated by electromagnetic interaction, we can write the helicity amplitude as follows
\begin{align}
	{\cal M}^{\lambda_{J/\psi}}_{\lambda_{\mu^-},\lambda_{\mu^+}}\left(J/\psi \to \mu^+\mu^-\right) = N_{J/\psi} \tilde{\varepsilon}_{J/\psi}^{\mu} \left(\lambda_{J/\psi}\right) \bar{u}\left(\lambda_{\mu^-}\right) \gamma_\mu v\left(\lambda_{\mu^+}\right),
\end{align}
where $N_{J/\psi} \equiv -8i\pi \alpha_{\rm EM} f_{J/\psi}/ \left( 3 m_{J/\psi}\right) $, $f_{J/\psi}$ is the decay constant of $J/\psi$ meson, $\alpha_{\rm EM}$ is the fine-structure constant. There are six nonzero helicity amplitudes as follows
\begin{align}
	{\cal M}^{1}_{\frac{1}{2},-\frac{1}{2}}&={\cal M}^{-1}_{-\frac{1}{2},\frac{1}{2}}=\frac{N_{J/\psi}m_{J/\psi}}{\sqrt{2}}(1+\cos\theta_{J/\psi}),
	\\
	{\cal M}^{-1}_{\frac{1}{2},-\frac{1}{2}}&={\cal M}^{1}_{-\frac{1}{2},\frac{1}{2}}=\frac{N_{J/\psi}m_{J/\psi}}{\sqrt{2}}(1-\cos\theta_{J/\psi}),
	\\
	{\cal M}^{0}_{\frac{1}{2},-\frac{1}{2}}&=-{\cal M}^{0}_{-\frac{1}{2},\frac{1}{2}}=N_{J/\psi}m_{J/\psi}\sin\theta_{J/\psi},
\end{align}
and one can obtain that the total decay rate is
\begin{align}
	\Gamma \left(J/\psi \to \mu^+\mu^-\right)=\frac{\left| N_{J/\psi}\right| ^2m_{J/\psi}}{12\pi}.
\end{align}

\subsection{Dimensionless factors}
\label{subsec:DF}

The dimensionless factors induced in the calculation of $V^* \to \tau^- (\to \pi^- \nu_\tau)\bar{\nu}_\tau$ are as follows~\cite{Hu:2020axt}
\begin{align}
S_t =& 2 \omega _{\pi } \kappa _{\tau }^2-\kappa _{\tau }^4-\kappa _{\pi }^2,
\label{eq:St}
\\
S_1 =& \frac{\kappa _{\tau }^2 }{8 \left(\omega _{\pi
	}^2 - \kappa _{\pi }^2\right)}\Big[\kappa _{\pi }^2 \left(-6 \omega _{\pi } \kappa
_{\tau }^2+3 \kappa _{\tau }^4+4 \omega _{\pi }^2+10 \omega _{\pi
}-5\right)
\nonumber\\
& + \left( 2 \omega _{\pi } - \kappa_{\tau}^2\right)  \left(2 \omega _{\pi }^2+2 \omega _{\pi
}-1\right) \kappa _{\tau }^2-3 \kappa _{\pi }^4+6 \left(1-2 \omega _{\pi
}\right) \omega _{\pi }^2\Big],
\\
S_2 =& \frac{\kappa _{\tau }^2 \left(\kappa _{\pi }^2-2 \omega _{\pi }+1\right)
	 \left(\omega _{\pi }-\kappa _{\tau
	}^2\right)}{\sqrt{\omega _{\pi }^2-\kappa _{\pi }^2}},
\\
S_3 =& \frac{\kappa _{\tau }^2}{8 \left(\omega _{\pi }^2 - \kappa _{\pi }^2\right)} \Big[\kappa _{\pi }^2 \left(-2 \omega _{\pi } \kappa
_{\tau }^2+\kappa _{\tau }^4+4 \omega _{\pi }^2-2 \omega _{\pi
}+1\right)
\nonumber\\
& + \left( \kappa _{\tau
}^2-2 \omega _{\pi }\right)  \left(2 \omega _{\pi }^2-6 \omega _{\pi }+3\right)
\kappa _{\tau }^2-\kappa _{\pi }^4+2 \left(1-2 \omega _{\pi }\right) \omega
_{\pi }^2\Big],
\\
S^T_1 =& \frac{1}{2 \left(\kappa
	_{\pi }^2-\omega _{\pi }^2\right)} \Big\{ \kappa _{\pi }^4 \left(2 \omega _{\pi } \kappa _{\tau }^2+5 \kappa
_{\tau }^4+2 \omega _{\pi }-3\right) +4
\omega _{\pi }^2 \kappa _{\tau }^2 \left[\left(3 \omega _{\pi }-1\right)
\kappa _{\tau }^2-\omega _{\pi }\right]
\nonumber\\
& +\kappa _{\pi }^2 \left[\left(-6 \omega
_{\pi }^2-10 \omega _{\pi }+3\right) \kappa _{\tau }^4+2 \left(3-2 \omega
_{\pi }\right) \omega _{\pi } \kappa _{\tau }^2+2 \omega _{\pi }^2\right]-\kappa _{\pi }^6 \Big\},
\\
S^T_2 =& \frac{4 \kappa _{\tau }^2 \left(\kappa _{\pi }^2-2 \omega _{\pi }+1\right)
	 \left(\kappa _{\pi }^2-\omega _{\pi } \kappa _{\tau
	}^2\right)}{\sqrt{\omega _{\pi }^2-\kappa _{\pi }^2}},
\\
S^T_3 =& \frac{1}{2
	\left(\omega _{\pi }^2 - \kappa _{\pi }^2\right)} \Big\{ \kappa _{\pi }^4 \left(-6 \omega _{\pi } \kappa _{\tau }^2+\kappa
_{\tau }^4-6 \omega _{\pi }+1\right) -4 \omega _{\pi }^2 \kappa _{\tau }^2 \left[\left(\omega _{\pi
}-1\right) \kappa _{\tau }^2+\omega _{\pi }\right]
\nonumber\\
&+\kappa _{\pi }^2 \left[\left(2 \omega
_{\pi }^2-2 \omega _{\pi }-1\right) \kappa _{\tau }^4+2 \omega _{\pi }
\left(6 \omega _{\pi }-1\right) \kappa _{\tau }^2+2 \omega _{\pi
}^2\right] +3 \kappa _{\pi }^6 \Big\},
\\
R_t =& \frac{\sqrt{2} \left(\omega _{\pi }-1\right) \kappa _{\tau } \left(2 \omega _{\pi } \kappa _{\tau }^2 - \kappa _{\tau }^4 - \kappa_{\pi }^2\right)}{\sqrt{\omega _{\pi }^2-\kappa _{\pi }^2}},
\\
R_t^T =& \frac{2 \sqrt{2} \left(\kappa _{\pi }^2-\omega _{\pi
	}\right) \left(-2 \omega _{\pi } \kappa _{\tau
	}^2+\kappa _{\tau }^4+\kappa _{\pi
	}^2\right)}{\sqrt{\omega _{\pi }^2-\kappa _{\pi
		}^2}},
\\
R_1 =& \frac{\kappa _{\tau }}{2 \left(\kappa _{\pi }^2-\omega _{\pi }^2\right)} \Big\{\kappa_{\pi }^2 \left[\left(\omega _{\pi }+2\right) \kappa _{\tau }^4+\left(4 \omega _{\pi }^2+8 \omega
_{\pi }-6\right) \kappa _{\tau }^2-4 \omega _{\pi }^2+\omega _{\pi }\right] 
\nonumber\\
&+\kappa _{\pi }^4 \left(-6 \kappa _{\tau }^2+\omega _{\pi }+2\right) +\omega _{\pi } \kappa
_{\tau }^2 \left[\left(1-4 \omega _{\pi }\right) \kappa _{\tau }^2-4 \left(\omega _{\pi }-1\right)
\omega _{\pi }\right]\Big\},
\\
R_2 =& \frac{2 \kappa _{\tau } \left(\kappa _{\tau }^4-\kappa _{\pi }^2\right) \left(\kappa _{\pi }^2-2 \omega
	_{\pi }+1\right)}{\sqrt{\omega _{\pi }^2-\kappa _{\pi }^2}},
\\
R_3 =& \frac{\kappa _{\tau } }{2 \left(\kappa _{\pi }^2-\omega _{\pi }^2\right)} \Big\{\kappa
_{\pi }^2 \left[\left(3 \omega _{\pi }-2\right) \kappa _{\tau }^4+\left(-4 \omega _{\pi }^2+8 \omega
_{\pi }-2\right) \kappa _{\tau }^2+\left(3-4 \omega _{\pi }\right) \omega _{\pi }\right]
\nonumber\\ 
&+\kappa _{\pi }^4 \left(-2 \kappa _{\tau }^2+3 \omega _{\pi }-2\right)+\omega _{\pi
} \kappa _{\tau }^2 \left[\left(3-4 \omega _{\pi }\right) \kappa _{\tau }^2+4 \left(\omega _{\pi
}-1\right) \omega _{\pi }\right]\Big\},
\label{eq:R3}
\end{align}
where the three dimensionless parameters are defined as
\begin{equation}
\label{eq:dimenless_paras}
\kappa_\tau \equiv  \frac{m_\tau}{\sqrt{q^2}},
\quad 
\kappa_\pi \equiv  \frac{m_\pi}{\sqrt{q^2}},
\quad
\omega_\pi \equiv  \frac{E_\pi}{\sqrt{q^2}}.
\end{equation}

In the limit $\kappa_\pi \to 0$, the $E_\pi$-integrated factors $\bar{S}_i(\bar{R}_i) \equiv \int S_i(R_i) dE_\pi$ are given, respectively, by
\begin{align}
\bar{S}_t  =& \frac{1}{4} \sqrt{q^2} \kappa _{\tau }^2 \left(\kappa _{\tau }^2-1\right)^2,
\\
\bar{S}_1 =& \frac{1}{16} \sqrt{q^2} \kappa _{\tau }^2 \left[\kappa _{\tau }^6-7 \kappa _{\tau }^4+3
   \kappa _{\tau }^2+8 \left(\kappa _{\tau }^4+\kappa _{\tau }^2\right) \log \left(\kappa
   _{\tau }\right)+3\right],
\\
\bar{S}_2 =& \frac{1}{4} \sqrt{q^2} \kappa _{\tau }^2 \left[-3 \kappa _{\tau }^4+2 \kappa _{\tau }^2+8
   \kappa _{\tau }^2 \log \left(\kappa _{\tau }\right)+1\right],
\\
\bar{S}_3 =& \frac{1}{16} \sqrt{q^2} \kappa _{\tau }^2 \left[-\kappa _{\tau }^6-21 \kappa _{\tau }^4+21
   \kappa _{\tau }^2+24 \left(\kappa _{\tau }^4+\kappa _{\tau }^2\right) \log \left(\kappa
   _{\tau }\right)+1\right],
\\
\bar{S}_1^T =& \frac{1}{4} \sqrt{q^2} \kappa _{\tau }^2 \left(\kappa _{\tau }^2-1\right)^2 \left(3 \kappa
   _{\tau }^2+1\right),
\\
\bar{S}_2^T =& -\sqrt{q^2} \kappa _{\tau }^4 \left(\kappa _{\tau }^2-1\right)^2,
\\
\bar{S}_3^T =& \frac{1}{4} \sqrt{q^2} \kappa _{\tau }^2 \left(\kappa _{\tau }^2-1\right)^3,
\\
\bar{R}_t =& \frac{1}{2 \sqrt{2}} \sqrt{q^2} \kappa _{\tau }^3 \left[\kappa _{\tau }^4+2 \kappa _{\tau }^2-8 \kappa
   _{\tau }^2 \log \left(\kappa _{\tau }\right)-3\right],
\\
\bar{R}_t^T =& \frac{1}{\sqrt{2}}\sqrt{q^2} \kappa _{\tau }^2 \left(\kappa _{\tau }^2-1\right)^2,
\\
\bar{R}_1 =& \frac{1}{4} \sqrt{q^2} \kappa _{\tau }^3 \left[-5 \kappa _{\tau }^4+8 \kappa _{\tau }^2+4
   \kappa _{\tau }^2 \log \left(\kappa _{\tau }\right)-3\right],
\\
\bar{R}_2 =& 2 \sqrt{q^2} \kappa _{\tau }^5 \left[\kappa _{\tau }^2-2 \log \left(\kappa _{\tau
   }\right)-1\right],
\\
\bar{R}_3 =& -\frac{3}{4} \sqrt{q^2} \kappa _{\tau }^3 \left[\kappa _{\tau }^4-4 \kappa _{\tau }^2 \log
   \left(\kappa _{\tau }\right)-1\right].
\end{align}
We provide the full analytical results for these factors electronically in the supplementary material.

\section{The detailed derivation of the dependence relations}
\label{sec:dep}

In this section we provide the detailed derivation of the dependence relations among the angular observables $\mathcal{I}_i$. It is useful to re-express $\mathrm{Im}[\mathcal{A}_{\parallel} \mathcal{A}_{0}^*]$, $\mathrm{Re}[\mathcal{A}_{\parallel}\mathcal{A}_{0}^*]$, 
$\mathrm{Re}[\mathcal{A}_{t}\mathcal{A}_{\perp}^*]$ and $\mathrm{Im}[\mathcal{A}_{t}
\mathcal{A}_{\parallel}^*]$ as
\begin{align}
\mathrm{Im}\left[\mathcal{A}_{\parallel} \mathcal{A}_{0}^*\right] =&\frac{1}{\left| \mathcal{A}_{\perp}\right|^2} \left\{\mathrm{Im}\left[\mathcal{A}_{\parallel} \mathcal{A}_{\perp}^*\right] \mathrm{Re}\left[\mathcal{A}_{0} \mathcal{A}_{\perp}^*\right] -\mathrm{Re}\left[\mathcal{A}_{\parallel} \mathcal{A}_{\perp}^*\right] \mathrm{Im}\left[\mathcal{A}_{0} \mathcal{A}_{\perp}^*\right]\right\},
\\
\mathrm{Re}\left[\mathcal{A}_{\parallel} \mathcal{A}_{0}^*\right] =&\frac{1}{\left| \mathcal{A}_{\perp}\right|^2} \left\{\mathrm{Re}\left[\mathcal{A}_{\parallel} \mathcal{A}_{\perp}^*\right] \mathrm{Re}\left[\mathcal{A}_{0} \mathcal{A}_{\perp}^*\right] +\mathrm{Im}\left[\mathcal{A}_{\parallel} \mathcal{A}_{\perp}^*\right] \mathrm{Im}\left[\mathcal{A}_{0} \mathcal{A}_{\perp}^*\right]\right\},
\\
\mathrm{Re}\left[\mathcal{A}_{t} \mathcal{A}_{\perp}^*\right] =&\frac{1}{\left| \mathcal{A}_{0}\right|^2} \left\{\mathrm{Re}\left[\mathcal{A}_{t} \mathcal{A}_{0}^*\right] \mathrm{Re}\left[\mathcal{A}_{0} \mathcal{A}_{\perp}^*\right] -\mathrm{Im}\left[\mathcal{A}_{t} \mathcal{A}_{0}^*\right] \mathrm{Im}\left[\mathcal{A}_{0} \mathcal{A}_{\perp}^*\right]\right\},
\\
\mathrm{Im}\left[\mathcal{A}_{t} \mathcal{A}_{\parallel}^*\right] =&\frac{1}{\left| \mathcal{A}_{0}\right|^2\left| \mathcal{A}_{\perp}\right|^2} \Big\{-\mathrm{Re}\left[\mathcal{A}_{t} \mathcal{A}_{0}^*\right] \mathrm{Re}\left[\mathcal{A}_{0} \mathcal{A}_{\perp}^*\right] \mathrm{Im}\left[\mathcal{A}_{\parallel} \mathcal{A}_{\perp}^*\right] \nonumber
\\
&+\mathrm{Re}\left[\mathcal{A}_{t} \mathcal{A}_{0}^*\right] \mathrm{Im}\left[\mathcal{A}_{0} \mathcal{A}_{\perp}^*\right] \mathrm{Re}\left[\mathcal{A}_{\parallel} \mathcal{A}_{\perp}^*\right]
+\mathrm{Im}\left[\mathcal{A}_{t} \mathcal{A}_{0}^*\right] \mathrm{Re}\left[\mathcal{A}_{0} \mathcal{A}_{\perp}^*\right] \mathrm{Re}\left[\mathcal{A}_{\parallel} \mathcal{A}_{\perp}^*\right]\nonumber
\\
&+\mathrm{Im}\left[\mathcal{A}_{t} \mathcal{A}_{0}^*\right] \mathrm{Im}\left[\mathcal{A}_{0} \mathcal{A}_{\perp}^*\right] \mathrm{Im}\left[\mathcal{A}_{\parallel} \mathcal{A}_{\perp}^*\right]
\Big\}.
\end{align}
Therefore, the twelve angular observables \eqref{eq:I1c_noT}--\eqref{eq:I7_noT} can be seen as functions of 4 real variables $\left| \mathcal{A}_{t}\right|^2$, $\left| \mathcal{A}_{\perp}\right|^2$, $\left| \mathcal{A}_{\parallel}\right|^2$, $\left| \mathcal{A}_{0}\right|^2$ and 3 complex variables $\mathcal{A}_{t} \mathcal{A}_{0}^*$, $\mathcal{A}_{0} \mathcal{A}_{\perp}^*$, $\mathcal{A}_{\parallel} \mathcal{A}_{\perp}^*$. The advantage of this view is that it implies the continuous symmetry \eqref{eq:conti_sym}. There are only seven independent real variables due to the following three relationships
\begin{align}
\mathrm{Re}\left[\mathcal{A}_{t} \mathcal{A}_{0}^*\right]^2 +\mathrm{Im}\left[\mathcal{A}_{t} \mathcal{A}_{0}^*\right]^2 
=& \left| \mathcal{A}_{t}\right|^2 \left| \mathcal{A}_{0}\right|^2,\label{eq:con1}
\\
\mathrm{Re}\left[\mathcal{A}_{0} \mathcal{A}_{\perp}^*\right]^2 +\mathrm{Im}\left[\mathcal{A}_{0} \mathcal{A}_{\perp}^*\right]^2 
=&\left| \mathcal{A}_{0}\right|^2 \left| \mathcal{A}_{\perp}\right|^2,\label{eq:con2}
\\
\mathrm{Re}\left[\mathcal{A}_{\parallel} \mathcal{A}_{\perp}^*\right]^2 +\mathrm{Im}\left[\mathcal{A}_{\parallel} \mathcal{A}_{\perp}^*\right]^2
=&\left| \mathcal{A}_{\parallel}\right|^2 \left| \mathcal{A}_{\perp}\right|^2. \label{eq:con3}
\end{align}

Inverting the eqs.~\eqref{eq:I1s_noT}--\eqref{eq:I5_noT}, one can rewrite the variables in terms of the angular observables $\mathcal{I}_i$ as follows
\begin{align}
\left|\mathcal{A}_{t}\right|^2
=&\frac{1}{S_t}\left[\mathcal{I}_{1s} + \frac{S_1-S_3}{2S_3} \mathcal{I}_{2s} - \frac{3S_1 - S_3}{4S_3} \mathcal{I}_{2c}\right],
\\
\left| \mathcal{A}_{0}\right|^2
=&\frac{1}{4S_3}\left(\mathcal{I}_{2c}-2\mathcal{I}_{2s}\right),
\\
\left| \mathcal{A}_{\perp}\right|^2
=&\frac{1}{2S_3} \left(\mathcal{I}_{2c}+\mathcal{I}_3\right),
\\
\left| \mathcal{A}_{\parallel}\right|^2 
=&\frac{1}{2S_3} \left(\mathcal{I}_{2c}-\mathcal{I}_3\right),
\\
\mathrm{Re}\left[\mathcal{A}_{t} \mathcal{A}_{0}^*\right] =&\frac{1}{2\sqrt{2}R_t}\left(\mathcal{I}_{6c}-2\mathcal{I}_{6s}\right),
\\
\mathrm{Im}\left[\mathcal{A}_{t} \mathcal{A}_{0}^*\right] 
=&\frac{S_3^2 \mathcal{I}_{6c} \beta_3  + S_2^2 \mathcal{I}_{9} \left(\mathcal{I}_{2c}-2 \mathcal{I}_{2s}\right) \beta_2}{2\sqrt{2} R_t S_2 S_3 \left(\mathcal{I}_{2c}+\mathcal{I}_{3}\right) \beta_2},
\\
\mathrm{Re}\left[\mathcal{A}_{0} \mathcal{A}_{\perp}^*\right]
=&\frac{1}{\sqrt{2}S_3}\mathcal{I}_{4},
\\
\mathrm{Im}\left[\mathcal{A}_{0} \mathcal{A}_{\perp}^*\right]
=&-\frac{S_2}{\sqrt{2} S_3^2 \mathcal{I}_{6c}} \beta_2,
\\
\mathrm{Re}\left[\mathcal{A}_{\parallel} \mathcal{A}_{\perp}^*\right]
=&\frac{1}{2S_2}\mathcal{I}_{6c},
\\
\mathrm{Im}\left[\mathcal{A}_{\parallel} \mathcal{A}_{\perp}^*\right]
=&\frac{1}{2S_3}\mathcal{I}_{9},
\end{align}
with $\beta_2$ and $\beta_3$ defined in eqs.~\eqref{eq:beta2} and \eqref{eq:beta3}, respectively. Substituting them into eqs.~\eqref{eq:I1c_noT}, \eqref{eq:I7_noT}, and \eqref{eq:con1}--\eqref{eq:con3}, we can obtain the final form of the 5 dependence relations \eqref{eq:relation1}--\eqref{eq:relation5}.

\bibliographystyle{JHEP}
\bibliography{ref}

\end{document}